\documentclass[a4paper,11pt]{article}

\usepackage{jheppub} % for details on the use of the package, please
\usepackage[T1]{fontenc} % if needed

\usepackage{subfigure}
\usepackage{comment}
\usepackage{hyperref}
\usepackage{amssymb}
\usepackage{graphicx}
\usepackage{amsmath}
\usepackage{tikz,fp}
\usepackage{tikz-cd}
\usetikzlibrary{arrows}
\usetikzlibrary{intersections}
\usetikzlibrary{shapes.geometric}
\usetikzlibrary{decorations.pathmorphing, patterns,shapes,fixedpointarithmetic}
\usetikzlibrary{decorations.markings}

\pgfdeclarepatternformonly{south west lines}{\pgfqpoint{-0pt}{-0pt}}{\pgfqpoint{3pt}{3pt}}{\pgfqpoint{3pt}{3pt}}{
        \pgfsetlinewidth{0.4pt}
        \pgfpathmoveto{\pgfqpoint{0pt}{0pt}}
        \pgfpathlineto{\pgfqpoint{3pt}{3pt}}
        \pgfpathmoveto{\pgfqpoint{2.8pt}{-.2pt}}
        \pgfpathlineto{\pgfqpoint{3.2pt}{.2pt}}
        \pgfpathmoveto{\pgfqpoint{-.2pt}{2.8pt}}
        \pgfpathlineto{\pgfqpoint{.2pt}{3.2pt}}
        \pgfusepath{stroke}}

\tikzset{
  mid arrow/.style={postaction={decorate,decoration={
        markings,
        mark=at position .575 with {\arrow{stealth}}
      }}},
  near arrow/.style={postaction={decorate,decoration={
        markings,
        mark=at position .275 with {\arrow{stealth}}
      }}},
  far arrow/.style={postaction={decorate,decoration={
        markings,
        mark=at position .800 with {\arrow{stealth}}
      }}},
  snake arrow/.style={fixed point arithmetic, decorate, decoration={snake,amplitude=2pt, segment length=11pt},postaction={decoration={markings,mark=at position 0.625 with {\arrow{stealth}}},decorate}},
}

\title{\boldmath Entanglement Entropy and its Quench Dynamics for Pure States of the Sachdev-Ye-Kitaev model}

\author[a,b]{Pengfei Zhang}
\affiliation[a]{Walter Burke Institute for Theoretical Physics, California Institute of Technology, Pasadena, CA 91125, USA}
\affiliation[b]{Institute for Quantum Information and Matter, California Institute of Technology, Pasadena, CA 91125, USA}
\emailAdd{pengfeizhang.physics@gmail.com}
\abstract{Sachdev-Ye-Kitaev (SYK) is a concrete solvable model with non-Fermi liquid behavior and maximal chaos. In this work, we study the entanglement R\'enyi entropy for the subsystems of the SYK model in the Kourkoulou-Maldacena states. We use the path-integral approach and take the saddle point approximation in the large-$N$ limit. We find a first-order transition exist when tuning the subsystem size for the $q=4$ case, while it is absent for the $q=2$ case. We further study the entanglement dynamics for such states under the real-time evolution for noninteracting, weakly interacting and strongly interacting SYK(-like) models. }

\begin{document} 
\maketitle
\flushbottom

\section{Introduction}

In recent years, the entanglement entropy and its dynamics in many-body systems have drawn a lot of attention. As an example, 
the entanglement entropy has been studied both theoretically \cite{Deutsch_1991,Srednicki_1994,calabrese2004entanglement,Garrison_2018} and experimentally \cite{islam2015measuring,li2017measuring} for interacting quantum systems that satisfy the eigenstate thermalization hypothesis (ETH). For a general energy eigenstate, it shows volume law scaling, in contrast to the area law scaling in the many-body localization (MBL) phase \cite{Bauer_2013,Lukin256}. The entanglement dynamics can also be related to the out-of-time-order correlators \cite{hosur2016chaos,fan2017out}, which characterize the scrambling of quantum information \cite{lashkari2013towards,kitaev2014hidden,Ho_2017,Mezei_2017}. Moreover, the recent resolution of the information paradox \cite{penington2019entanglement,almheiri2019entropy,almheiri2019page,almheiri2019islands,almheiri2019replica,penington2019replica} is directly from the refined understanding of the Ryu-Takayanagi formula \cite{Ryu:2006bv,Ryu:2006ef,Lewkowycz:2013nqa} for computing the entanglement entropy in holographic systems.

Unfortunately, the calculation of the entanglement entropy for many-body systems is usually hard. Toy models where entanglement entropy can be computed efficiently are of especial interest. One strategy is to construct random unitary dynamics \cite{hayden2016holographic,Nahum_2017,von2018operator}. 
Here, we consider an alternative route by studying a specific solvable model named the Sachdev-Ye-Kitaev model \cite{sachdev1993gapless,kitaev2014hidden,maldacena2016remarks,kitaev2018soft}, which describes $N$ Majorana modes with infinite range $q-$body interaction. For $q\geq 4$, it is known as a non-Fermi liquid without quasiparticles which shows maximal chaotic behavior. In the low-energy limit, the system is described in terms of reparametrization modes, with an effective Schwarzian action \cite{kitaev2018soft,maldacena2016remarks,kitaev2014hidden}. The same action also shows up for the Jackiw-Teitelboim gravity in 2D \cite{kitaev2018soft,maldacena2016conformal}. 

Previously, there are studies of the SYK model from the entropy perspective. Assuming the system satisfies ETH \cite{Deutsch_1991,Srednicki_1994,hunter2018thermalization,haque2019eigenstate,sonner2017eigenstate}, the entanglement entropy can be related to the thermal entropy with an effective temperature depending on the system size \cite{huang2019eigenstate}. An analytical approximation has also been derived based on the many-body spectrum \cite{garcia2017analytical}. Numerically, the subsystem entropy has been studied in \cite{liu2018quantum} by exact diagonalization for the ground state, and in \cite{new,haldar2020renyi} by path-integral approach for thermal ensembles. There are also studies for two copies of (coupled) SYK models prepared in the thermofield double state, which purifies the thermal density matrix \cite{gu2017spread,penington2019replica,Yiming}. There are also studies for random hopping model, which is directly related to the SYK$_2$ case \cite{magan2016black,magan2016random}. However, the large-$N$ microscopic entanglement entropy and its dynamics for pure states of a single SYK model are still unknown \footnote{There are related dicussions of entanglement entropy in \cite{magan2017finetti}.}. 

In this work, we establish the formulation for computing R\'enyi entanglement entropy for the Kourkoulou-Maldacena pure states in SYK-like models \cite{kourkoulou2017pure}. The paper is organized as follows: In section \ref{sec2}, we give a brief review of the SYK model and the Kourkoulou-Maldacena states. We then derive the path-integral representation of R\'enyi entanglement entropy for such pure states in section \ref{sec3}. The numerical results are presented in section \ref{sec4}. By varying the subsystem size, we find a first-order transition of the entanglement entropy for the SYK$_4$ model, which leads to the singularity of the Page curve at half system size \cite{page1993average}. This is qualitatively different from the $q=2$ case, where the entanglement entropy changes analytically when varying the subsystem size. We further study the exact dynamics of the entanglement entropy for the SYK$_4$ non-Fermi liquid, with a comparison to noninteracting or weakly interacting SYK$_2$ Fermi liquids in section \ref{sec5}. Finally, we summarize our results in \ref{sec6}.

\section{SYK model and Kourkoulou-Maldacena states}\label{sec2}

The Hamiltonian of SYK$_q$ model \cite{kitaev2014hidden,maldacena2016remarks} reads:
\begin{equation}
H=\frac{1}{q!}\sum_{i_1i_2...i_q}i^{q/2}J^{(q)}_{i_1i_2...i_q}\chi_{i_1}\chi_{i_2}...\chi_{i_q}.\label{H0}
\end{equation}
Here $q$ is an even integer and $i=1,2...N$ labels different Majorana modes. We take the convention that $\{\chi_i,\chi_j\}=\delta_{ij}$. $J^{(q)}_{i_1i_2...i_q}$ are independent Gaussian variables with:
\begin{equation}
\overline{|J^{(q)}_{i_1i_2...i_q}|}=0,\ \ \ \ \ \ \overline{|J^{(q)}_{i_1i_2...i_q}|^2}=\frac{(q-1)!J^2}{N^{q-1}}=\frac{2^{q-1}(q-1)!\mathcal{J}^2}{qN^{q-1}}. \label{randomness}
\end{equation}
Here $\mathcal{J}$ is taken to be a constant in the large-$q$ limit.

For a thermal ensemble, to the leading order of $1/N$, the two-point correlator $G_{\text{th}} (\tau)=\left<\mathcal{T}_\tau \chi_i(\tau)\chi_i(0)\right>_\beta$ satisfies the self-consistent equation:
\begin{equation}
G^{-1}_\text{th}(i\omega_n)=-i\omega_n-\Sigma_\text{th}(i\omega_n),\ \ \ \ \ \ \Sigma_\text{th}(\tau)=\begin{tikzpicture}[baseline={([yshift=-6pt]current bounding box.center)}, scale=1.1]
\draw[thick] (-22pt,0pt) -- (-13pt,0pt);
\draw[dashed,thick] (-13pt,0pt)..controls (-8pt,18pt) and (8pt,18pt)..(13pt,0pt);
\draw[thick] (-13pt,0pt)..controls (-8pt,10pt) and (8pt,10pt)..(13pt,0pt);
\draw[thick] (-13pt,0pt)..controls (-8pt,-10pt) and (8pt,-10pt)..(13pt,0pt);
\draw[thick] (-13pt,0pt)..controls (-8pt,5pt) and (8pt,5pt)..(13pt,0pt);
\draw[thick] (13pt,0pt) -- (22pt,0pt);
\draw (0,0) node{$.$};
\draw (0,-2pt) node{$.$};
\draw (0,-4pt) node{$.$};
\end{tikzpicture}=J^2G_\text{th}^{q-1}(\tau),\label{SDTH}
\end{equation}
where the self-energy is given by melon diagrams. By solving the Schwinger-Dyson equation, the model is found to be a Fermi liquid with finite spectral function near $\omega\sim0$ for $q=2$. On contrary, for $q\geq4$, the model has divergent spectral function $\rho(\omega)\sim \omega^{2/q-1}$, which is known as a non-Fermi liquid. Further study shows it has maximal chaos \cite{kitaev2014hidden,maldacena2016remarks} and satisfies the ETH \cite{hunter2018thermalization,haque2019eigenstate,sonner2017eigenstate}. 

Considering the SYK system in some eigenstate $|E\rangle$ with energy $E=N\epsilon$, the entropy $\mathcal{S}_A$ of the a subsystem $A$ containing $M=\lambda N$ ($\lambda<1/2$) Majorana fermions is argued to be \cite{huang2019eigenstate}:
\begin{equation}
\mathcal{S}_A=Ms_{\text{th}}(\lambda^{\frac{q-1}{2}}\epsilon), \label{ETH}
\end{equation}
for $q\geq 4$. Here $s_{\text{th}}(x)$ is the thermal entropy density in the micro-canonical ensemble with energy density $x$. A similar statement when the total system is prepared in a thermal ensemble has been tested in \cite{new}. Approximately, we have $s_{\text{th}}(x)\approx (\log(2)/2 -  \arcsin (x/\epsilon_0)/q^2 )$ with $\epsilon_0$ being the energy density of the ground state \cite{garcia2017analytical}. On the other hand, for $q=2$, the ground state entanglement entropy can be calculated analytically \cite{liu2018quantum,new}. 

In this work, we focus on a specific class of pure states of the SYK model \cite{kourkoulou2017pure}. These states are now known as the Kourkoulou-Maldacena (KM) states. To construct them, we first pair Majorana fermions as $c_j=\frac{\chi_{2j-1}+i\chi_{2j}}{2}$ with $j=1,2...N/2$ \footnote{Since there is a permutation symmetry for different modes $i$, this choice is general.}. Then (unnormalized) KM states are given by:
\begin{equation}
|\text{KM}(\{s\},\beta)\rangle=e^{-\frac{\beta H}{2}}|\{s\}\rangle,\ \ \ \ \ \ (2n_j-1)|\{s\}\rangle=s_j|\{s\}\rangle.
\end{equation}
Here $n_j=c^\dagger_j c_j$ and $s_j\in \{\pm1\}$. We could always redefine $\chi_{2j}\rightarrow s_j \chi_{2j}$ to set $s_j=1$ for all $j$. As result, all single states are equivalent after averaging over the ensemble of random interaction. We will focus on $\{s\}=\{1\}$ for most parts of the manuscript. Moreover, for simplicity, from now on we keep the $\beta= 1/T$ dependence of $\left|\text{KM}\right>$ implicit. 

The hallmark of these states is that to the leading order of $1/N$, the correlation functions of $\chi_i$ can be related to thermal correlators \cite{kourkoulou2017pure}, under the assumption of the disorder replica diagonal \cite{fu2016numerical,gur2018does,kitaev2018soft,gu2019notes}. As an example, two-point functions $$G_{ij}(\tau,\tau')=\frac{1}{Z_{\text{KM}}}\left<\{1\}\right|\mathcal{T}_\tau e^{-\int_0^\beta d\tau H}\chi_{i}(\tau)\chi_{i}(\tau')\left|\{1\}\right>,$$ with $Z_{\text{KM}}\equiv \langle \text{KM}| \text{KM} \rangle \equiv e^{-I_{\text{KM}}}$, can be expressed in terms of the thermal Green's function $G_{\text{th}} (\tau)$. Explicitly, all non-zero components are
\begin{equation}
\begin{aligned}
&G_{ii}(\tau,\tau')=G_{\text{th}}(\tau-\tau'), \ \ \ \ \ \ G_{2j-1,2j}(\tau,\tau')=-i2G_{\text{th}}(\tau)G_{\text{th}}(\tau').\label{Gij}
\end{aligned}
\end{equation}
Here we have $\tau, \tau' \in[0,\beta]$. This shows that the diagonal component $G_{ii}$ take the same form as a thermal Green's function, while the off-diagonal part $G_{2j-1,2j}(\beta/2,\beta/2)$ characterize the deviation from a thermalized state (at the two-point function level). For $\beta J\rightarrow \infty$, $|\text{KM}\rangle$ selects one state from the ground state sector of the SYK model, and $G_{2j-1,2j}(\beta/2,\beta/2)\rightarrow 0$.

We are mainly interested in the R\'enyi entanglement entropy of such pure states. For such states, we define subsystem A consisting of $M/2$ complex fermions \footnote{Note that although the KM pure states are expected to be dual to an AdS$_2$ geometry with a brane, there is no index $i$ degree of freedom, and consequently no direct analogy of this entanglement entropy in the gravity picture.}. The reduced density matrix $\rho_A=\frac{1}{Z_\text{KM}}\text{tr}_{B} |\text{KM} \rangle \langle \text{KM}|\equiv \tilde{\rho}_A/Z_{\text{KM}}$ is given by tracing out its complimentary $B$. The $n-$th R\'enyi entropy is then given by 
\begin{equation}
\mathcal{S}_A^{(n)}=\frac{1}{1-n}\log(\text{tr}_A\rho_A^n),
\end{equation}
with $\mathcal{S}_A\equiv\mathcal{S}_A^{(1)}$ being the Von Neumann entropy. Since the full system is in a pure state, we expect $\mathcal{S}_A^{(n)}$ being symmetric under a reflection along $\lambda=1/2$. For $\beta=0$, $|\text{KM}\rangle$ is a product state, and we have $\mathcal{S}_A^{(n)}=0$. On the contrary, if we take $\beta J \rightarrow \infty$, and we expect $\mathcal{S}_A^{(n)}$ follow a Page curve \cite{page1993average} with energy density depending on subsystem size \eqref{ETH}. 

\section{Path-integral for pure-state entanglement entropy}\label{sec3}
In this section, we derive the path-integral representation of $\mathcal{S}_A^{(n)}$ for KM pure states. We begin with a warm up by computing the normalization factor $Z_\text{KM}$ in section \ref{sec31}, which is also needed then computing the entanglement entropy. The path integral formula for computing $\mathcal{S}_A^{(n)}$ is then derived in \ref{sec32}.

\subsection{A warm up: $\left<\text{KM}|\text{KM}\right>$}\label{sec31}
Let us first consider the path-integral representation of $Z_\text{KM}=e^{-I_{\text{KM}}}=\left<\text{KM}|\text{KM}\right>$. Note that this is in fact not essential, since $Z_{\text{KM}}$ can be directly related to the the thermal partition function $Z=\text{tr}\ e^{-\beta H}$ \cite{new}. However, the trick developed in this subsection is useful when computing the entanglement entropy.

The state $|\text{KM}\rangle$ is given by an imaginary-time evolution of a initial state $|\{1\}\rangle$. The graphic representation is:
\begin{equation}
|\text{KM}\rangle=e^{-\frac{\beta H}{2}}|\{1\}\rangle=\begin{tikzpicture}[thick,scale = 0.5,baseline={([yshift=0pt]current bounding box.center)}]
 \draw (0.7,0.8) arc(0:-180:0.5 and 0.5);

 \draw  (0.7,0.8)  --  (0.7,2.8);
 \draw  (-0.3,0.8)  --  (-0.3,2.8);

 \draw[dotted]  (-0.3,0.8)  --  (0.7,0.8);
 \draw[dotted]  (-0.3,1.4)  --  (0.7,1.4);
 \draw[dotted]  (-0.3,2)  --  (0.7,2);
 \draw[dotted]  (-0.3,2.6)  --  (0.7,2.6);

 \filldraw  (0.2,0.3) circle (1.5pt) node[left]{\scriptsize $ $};

 \draw (1,-0.1) node{\scriptsize $|\{s\}\rangle$};
 \draw (1,0.6) node{\scriptsize $0$};
 \draw (1,2.8) node{\scriptsize $\frac{\beta}{2}$};
 \draw (-0.6,0.6) node{\scriptsize $0$};
 \draw (-0.6,2.8) node{\scriptsize $\frac{\beta}{2}$};

 \draw (1.4,1.7) node{\scriptsize $\chi_2$};
 \draw (-1,1.7) node{\scriptsize $\chi_1$};
\end{tikzpicture}
\end{equation}
Here we have explicitly separated out fermions with odd/even indices: $\chi_{1/2}$ represents Majorana fermions $\chi_{2j-1}/\chi_{2j}$ with odd/even indices. The solid lines denote the imaginary-time evolution and the dotted lines represent interactions between fermions. Two points connected by the dotted line are at the same imaginary time. The black dots represent the boundary condition $c_j|\{1\}\rangle=0$, or in terms of Majorana fermions $\left(\chi_{2j-1}+i\chi_{2j}\right)|\{1\}\rangle=0$. Similarly, the normalization $Z_{\text{KM}}$ is given by 
\begin{equation}
Z_{\text{KM}}=e^{-I_{\text{KM}}}=\langle \{ 1 \}|e^{-\beta H}|\{1\}\rangle=\begin{tikzpicture}[thick,scale = 0.45,baseline={([yshift=0pt]current bounding box.center)}]
 \draw (0.7,0.8) arc(0:-180:0.5 and 0.5);
 \draw (0.7,3.8) arc(0:180:0.5 and 0.5);

 \draw  (0.7,0.8)  --  (0.7,3.8);
 \draw  (-0.3,0.8)  --  (-0.3,3.8);

 \draw[dotted]  (-0.3,0.8)  --  (0.7,0.8);
 \draw[dotted]  (-0.3,1.4)  --  (0.7,1.4);
 \draw[dotted]  (-0.3,2)  --  (0.7,2);
 \draw[dotted]  (-0.3,2.6)  --  (0.7,2.6);
 \draw[dotted]  (-0.3,3.2)  --  (0.7,3.2);
 \draw[dotted]  (-0.3,3.8)  --  (0.7,3.8);

 \filldraw  (0.2,0.3) circle (1.5pt) node[left]{\scriptsize $ $};
 \filldraw  (0.2,4.3) circle (1.5pt) node[left]{\scriptsize $ $};

 \draw (1,-0.1) node{\scriptsize $|\{s\}\rangle$};
 \draw (0,4.7) node{\scriptsize $\langle \{s\}|$};
 \draw (1,0.6) node{\scriptsize $0$};
 \draw (1,3.9) node{\scriptsize $\beta$};
  \draw (-0.6,0.6) node{\scriptsize $0$};
 \draw (-0.6,3.9) node{\scriptsize $\beta$};

 \draw (1.4,2.2) node{\scriptsize $\chi_2$};
 \draw (-1,2.2) node{\scriptsize $\chi_1$};
\end{tikzpicture}
\end{equation}
Here we have another boundary condition $\langle\{1\}|\left(\chi_{2j-1}-i\chi_{2j}\right)=0$ at imaginary time $\beta$. The path-integral representation of $Z_{\text{KM}}$ is then
\begin{equation}
\begin{aligned}
e^{-I_{\text{KM}}}&=\int_{\text{b.c.}}\mathcal{D}\chi_{i}(\tau) \exp(-S_{\text{KM}}[\chi_{i}]),\\
S_{\text{KM}}&= \int_0^{\beta}d\tau \left(\frac{1}{2}\sum_i\chi_{i}\partial_\tau\chi_{i}+\frac{1}{q!}\sum_{i_1i_2...i_q}i^{q/2}J^{(q)}_{i_1i_2...i_q}\chi_{i_1}\chi_{i_2}...\chi_{i_q}\right).
\end{aligned}
\end{equation}
Here b.c. indicates the the boundary condition at $\tau=0$ and $\tau=\beta$:
\begin{equation}
\chi_{2j-1}(0)=-i\chi_{2j}(0),\ \ \ \ \ \ \chi_{2j-1}(\beta)=i\chi_{2j}(\beta).\label{bdy}
\end{equation} 

We further take the disorder average of random interaction $J^{(q)}_{i_1i_2...i_q}$. As for the thermal ensemble, we expect the replica diagonal assumption works well in the large-$N$ limit and we could neglect the difference between $\exp(-\overline{I_{\text{KM}}})$ and $\overline{\exp(-I_{\text{KM}})}$. Consequently, we keep the disorder average implicitly from now on. 

To proceed, we use the standard trick by introducing bilocal fields $G$ and $\Sigma$ \cite{maldacena2016remarks}. Since fields with even or odd indices are in-equivalent, we should define two sets of fields $G_{11/22}$ and $\Sigma_{11/22}$. The definition of $G_{11}$ and $G_{22}$ is
\begin{equation}
G_{11}(\tau,\tau')=\frac{2}{N}\sum_{j}\chi_{2j-1}(\tau)\chi_{2j-1}(\tau'),\ \ \ \ \ \ G_{22}(\tau,\tau')=\frac{2}{N}\sum_{j}\chi_{2j}(\tau)\chi_{2j}(\tau').
\end{equation}
$\Sigma_{11}$ and $\Sigma_{22}$ are introduced in order to impose the relation between $G$ and $\chi\chi$:
\begin{equation}
\begin{aligned}
\delta\left(G_{11}-\frac{2}{N}\sum_{i\in\text{odd}}\chi_{i}\chi_{i}\right)&=\int \mathcal{D}\Sigma_{22}\ e^{\frac{1}{2}\int d\tau d\tau'\Sigma_{11}(\tau,\tau')\left(\sum_{\text{odd}} \chi_{i}(\tau)\chi_{i}(\tau')-\frac{N}{2}G_{11}(\tau,\tau')\right)},\\
\delta\left(G_{22}-\frac{2}{N}\sum_{i\in\text{even}}\chi_{i}\chi_{i}\right)&=\int \mathcal{D}\Sigma_{22}\ e^{\frac{1}{2}\int d\tau d\tau'\Sigma_{22}(\tau,\tau')\left(\sum_\text{even} \chi_{i}(\tau)\chi_{i}(\tau')-\frac{N}{2}G_{22}(\tau,\tau')\right)}.
\end{aligned}
\end{equation} 
Then by integrating out the Majorana fields, we find
\begin{equation}
e^{-I_{\text{KM}}}=\int \mathcal{D}G_{11} \mathcal{D}G_{22} \mathcal{D}\Sigma_{11} \mathcal{D}\Sigma_{22} \exp(-S_{\text{KM}}^{\text{eff}}[G,\Sigma]).
\end{equation}
Here the effective $G$-$\Sigma$ action $S_{\text{KM}}$ is given by \footnote{There could be additional boundary terms. Nevertheless, they cancel out when computing the entanglement entropy.}
\begin{equation}
\begin{aligned}
\frac{S_{\text{KM}}^{\text{eff}}}{N}=&-\frac{1}{4}\log \underset{\text{b.c.}}{\det} 
\begin{pmatrix}
\partial_\tau-\Sigma_{11}&0\\
0&\partial_\tau-\Sigma_{22}
\end{pmatrix}
-\frac{J^2}{2q}\int d\tau d\tau' \left(\frac{G_{11}(\tau,\tau')+G_{22}(\tau,\tau')}{2}\right)^q \\
&+\frac{1}{4}\int d\tau d\tau'G_{11}(\tau,\tau')\Sigma_{11}(\tau,\tau')+\frac{1}{4}\int d\tau d\tau'G_{22}(\tau,\tau')\Sigma_{22}(\tau,\tau').\\
\end{aligned}\label{actionKM0}
\end{equation} 
Here $\text{b.c.}$ denotes the boundary condition \ref{bdy}. Note that although the self-energy is blocked diagonal, the boundary condition would mix modes with even/odd indices. In the large-$N$ limit, we take the saddle point of \eqref{actionKM0}. The saddle point equation reads 
\begin{equation}
\begin{aligned}
\begin{pmatrix}
G_{11}&G_{12}\\
G_{21}&G_{22}
\end{pmatrix}&=
\begin{pmatrix}
\partial_\tau-\Sigma_{11}&0\\
0&\partial_\tau-\Sigma_{22}
\end{pmatrix}_\text{b.c.}^{-1},\\
\Sigma_{11}(\tau,\tau')=\Sigma_{22}(\tau,&\tau')=J^2\left(\frac{G_{11}(\tau,\tau')+G_{22}(\tau,\tau')}{2}\right)^{q-1}.
\end{aligned}\label{SDKM0}
\end{equation}
Here we have also introduced $$G_{12}(\tau,\tau')=\frac{2}{N}\sum_j\left<\mathcal{T}_\tau\chi_{2j-1}(\tau)\chi_{2j}(\tau')\right>, \ \ \ \ \ \ G_{21}(\tau,\tau')=\frac{2}{N}\sum_j\left<\mathcal{T}_\tau\chi_{2j}(\tau)\chi_{2j-1}(\tau')\right>,$$ for completeness. In terms of bilocal fields $G_{ab}$ with $a,b\in \{1,2\}$, the boundary condition \eqref{bdy} becomes 
\begin{equation}
\begin{aligned}
&G_{1a}(0,\tau)=-iG_{2a}(0,\tau),\ \ \ \ \ \ G_{1a}(\beta,\tau)=iG_{2a}(\beta,\tau),\\
&G_{a1}(\tau,0)=-iG_{a2}(\tau,0),\ \ \ \ \ \ G_{a1}(\tau,\beta)=iG_{a2}(\tau,\beta).
\end{aligned}\label{bdy2}
\end{equation}

Solving the equation \eqref{SDKM0} with boundary condition \eqref{bdy2}, and substituting the solution into \eqref{actionKM0} already gives the on-shell action $I_{\text{KM}}$ and thus $Z_{\text{KM}}$. However, it is more convenient to introduce a different parametrization of the contour. The key observation is that if we define a Majorana field $\chi_j(s)$ with parameter $s\in[0,2\beta)$:
\begin{equation}
\begin{tikzpicture}[thick,scale = 0.45,baseline={([yshift=-4pt]current bounding box.center)}]
 \draw (0.7,0.8) arc(0:-180:0.5 and 0.5);
 \draw (0.7,3.8) arc(0:180:0.5 and 0.5);

 \draw  (0.7,0.8)  --  (0.7,3.8);
 \draw  (-0.3,0.8)  --  (-0.3,3.8);

 \draw[dotted]  (-0.3,0.8)  --  (0.7,0.8);
 \draw[dotted]  (-0.3,1.4)  --  (0.7,1.4);
 \draw[dotted]  (-0.3,2)  --  (0.7,2);
 \draw[dotted]  (-0.3,2.6)  --  (0.7,2.6);
 \draw[dotted]  (-0.3,3.2)  --  (0.7,3.2);
 \draw[dotted]  (-0.3,3.8)  --  (0.7,3.8);

 \filldraw  (0.2,0.3) circle (1.5pt) node[left]{\scriptsize $ $};
 \filldraw  (0.2,4.3) circle (1.5pt) node[left]{\scriptsize $ $};

 \draw (0.8,0) node{\scriptsize $0$};
 \draw (-0.4,0) node{\scriptsize $2\beta$};
 \draw (0,4.6) node{\scriptsize $\beta$};

 \draw (-1.2,2.2) node{\scriptsize $\chi(s)$};
\end{tikzpicture}:\ \ \ \ \ \ \ \chi_j(s)=\begin{cases}
\chi_{2j}(s)\ \ \ \ \ \ \ \ \ \ \ \ \ \ \ \ \text{for}\ s\in[0,\beta)\\
-i\chi_{2j-1}(2\beta-s)\ \ \ \text{for}\ s\in[\beta,2\beta)
\end{cases},\label{defchi0}
\end{equation}
here $j=1,2...N/2$, the boundary condition \eqref{bdy} becomes the traditional continuous and anti-periodic boundary condition $\chi_j(2\beta^-)=-\chi_j(0^+)$, as for a thermal ensemble. The Green's function $G(s,s')=\left<\mathcal{T}_{\mathcal{C}} \chi_j(s)\chi_j(s')\right>$ is then given by
\begin{equation}
G(s,s')=
\begin{pmatrix}
G_{22}(s,s')&-iG_{21}(s,2\beta-s')\\
-iG_{12}(2\beta-s,s')&-G_{11}(2\beta-s,2\beta-s')
\end{pmatrix}.
\end{equation}
The self-consistent equation for $G(s,s')$ is then
\begin{equation}G(s,s')=(\partial_s-\Sigma)^{-1}(s,s')\equiv 
\begin{pmatrix}
\partial_s-\Sigma_{22}(s,s')& 0\\
0& \partial_s+\Sigma_{11}(2\beta-s,2\beta-s')
\end{pmatrix}^{-1}.
\end{equation}
Moreover, both the action \eqref{actionKM0} and boundary condition \eqref{bdy} are invariant under $\chi_{2j-1}\rightarrow \chi_{2j}$ and $\chi_{2j}\rightarrow -\chi_{2j-1}$. As a result, we have $G_{11}(\tau,\tau')=G_{22}(\tau,\tau')$. Instead of \eqref{SDKM0}, we can then use  
\begin{equation}
\begin{aligned}
\Sigma(s,s')&=J^2G^{q-1}(s,s')P(s,s'),\\
P(s,s')&=\left[\theta(\beta-s)\theta(\beta-s')+\theta(s-\beta)\theta(s'-\beta)\right].
\end{aligned}\label{SDKM0s}
\end{equation}
Here $P(s,s')$ is a projector. We have $P(s,s')=1$ if both $\chi(s)$ and $\chi(s')$ represents the same field ($\chi_1$ or $\chi_2$) and otherwise zero. This definition for $P(s,s')$ is more general if we choose a different $s=0$ point on the contour \eqref{defchi0}. Note that comparing to a thermofield double state with inverse temperature $2\beta$ and $N/2$ fermions, the main difference is the presence of $P(s,s')$, which breaks the time translational invariance.

We can then choose to solve the equation \eqref{SDKM0s} and $G=(\partial_s-\Sigma)^{-1}$ self-consistently. Moreover, we could also express the on-shell action in terms of $G(s,s')$ and $\Sigma(s,s')$: 
\begin{equation}
\begin{aligned}
\frac{I_{\text{KM}}}{N}=&\frac{1}{4}\log \det G+\frac{q-1}{4q}\int ds ds'G(s,s')\Sigma(s,s').\\
\end{aligned}\label{actionKM0}
\end{equation}
This gives an alternative route to compute $Z_{\text{KM}}$.

\subsection{Computing $\mathcal{S}_A^{(n)}$ }\label{sec32}

Having illustrated the trick of parameterizing the contour by $s$, we consider the path-integral representation of $\mathcal{S}_A^{(n)}$ in this subsection. 

To compute $\mathcal{S}_A^{(2)}$, we first consider the path-integral representation of $|\text{KM} \rangle \langle \text{KM}|$. Separating out the modes in system $A$ and $B$, a graphic representation is

\begin{equation}
|\text{KM} \rangle \langle \text{KM}|\ =\begin{tikzpicture}[thick,scale = 0.45,baseline={([yshift=-2pt]current bounding box.center)}]
 \draw[blue] (0.7,0.8) arc(0:-180:0.5 and 0.5);
 \draw[blue] (0.7,-0.8) arc(0:180:0.5 and 0.5);
 \draw[red] (0.3,0.8) arc(0:-180:0.5 and 0.5);
 \draw[red] (0.3,-0.8) arc(0:180:0.5 and 0.5);

 \draw[blue]  (0.7,0.8)  --  (0.7,2.8);
 \draw[blue]  (-0.3,0.8)  --  (-0.3,2.8);
 \draw[red]  (-0.7,0.8)  --  (-0.7,2.8);
 \draw[red]  (0.3,0.8)  --  (0.3,2.8);
 \draw[blue]  (0.7,-0.8)  --  (0.7,-2.8);
 \draw[blue]  (-0.3,-0.8)  --  (-0.3,-2.8);
 \draw[red]  (-0.7,-0.8)  --  (-0.7,-2.8);
 \draw[red]  (0.3,-0.8)  --  (0.3,-2.8);

 \draw[dotted]  (-0.7,0.8)  --  (0.7,0.8);
 \draw[dotted]  (-0.7,1.4)  --  (0.7,1.4);
 \draw[dotted]  (-0.7,2)  --  (0.7,2);
 \draw[dotted]  (-0.7,2.6)  --  (0.7,2.6);

  \draw[dotted]  (-0.7,-0.8)  --  (0.7,-0.8);
 \draw[dotted]  (-0.7,-1.4)  --  (0.7,-1.4);
 \draw[dotted]  (-0.7,-2)  --  (0.7,-2);
 \draw[dotted]  (-0.7,-2.6)  --  (0.7,-2.6);

 \filldraw  (0.2,0.3) circle (1.5pt) node[left]{\scriptsize $ $};
 \filldraw  (-0.2,0.3) circle (1.5pt) node[left]{\scriptsize $ $};
 \filldraw  (0.2,-0.3) circle (1.5pt) node[left]{\scriptsize $ $};
 \filldraw  (-0.2,-0.3) circle (1.5pt) node[left]{\scriptsize $ $};

 \draw[red] (-0.7,0) node{\scriptsize $A$};
 \draw[blue] (0.7,0) node{\scriptsize $B$};
 \draw[red] (-0.7,3.2) node{\scriptsize $\chi_{1}^A$};
 \draw[red] (0.3,3.2) node{\scriptsize $\chi_{2}^A$};
 \draw[blue] (0.7,-3.2) node{\scriptsize $\chi_{2}^B$};
 \draw[blue] (-0.3,-3.2) node{\scriptsize $\chi_{1}^B$};
 \draw (-1.7,1.5) node{\scriptsize $|\text{KM}\rangle$};
 \draw (-1.7,-1.5) node{\scriptsize $\langle\text{KM}|$};
 \draw (1.,0.7) node{\scriptsize $0$};
 \draw (1.,2.6) node{\scriptsize $\frac{\beta}{2}$};
\end{tikzpicture},
\end{equation}
here the red/blue solid line represents the contour for subsystem $A/B$. $\chi_{1/2}^{S}$ represents Majorana fermions in subsystem $S$ with odd/even indices, with $S\in\{A,B\}$. 

The unnormalized density matrix $\tilde{\rho}_A=\text{tr}_B |\text{KM} \rangle \langle \text{KM}|$ is then given by tracing out the $B$ subsystem or, graphically, by connecting the (blue) contours of $B$. $\text{tr}_A \tilde{\rho}_A^n$ can then be computed by sewing $n$ copies of $\tilde{\rho}_A$. To be concrete, in this work we focus on the $n=2$ case. The corresponding contour is then given by
\begin{equation}
\text{tr}_A \tilde{\rho}_A^2=\text{tr}_A (\text{tr}_B |\text{KM} \rangle \langle \text{KM}|)^2=\begin{tikzpicture}[thick,scale = 0.45,baseline={([yshift=-11pt]current bounding box.center)}]
 \draw[dashed,thin]  (-3.5,0)  --  (3.5,0);
 \draw[red] (-2,-1.5) arc(0:-180:0.5 and 0.5);
 \draw[red] (-2,1.5) arc(0:180:0.5 and 0.5);
 \draw[red] (2,-1.5) arc(-180:0:0.5 and 0.5);
 \draw[red] (2,1.5) arc(180:0:0.5 and 0.5);
 \draw[blue] (-1.5,2) arc(-90:-270:0.5 and 0.5);
 \draw[blue] (-1.5,-2) arc(90:270:0.5 and 0.5);
 \draw[blue] (1.5,2) arc(-90:90:0.5 and 0.5);
 \draw[blue] (1.5,-2) arc(90:-90:0.5 and 0.5);

\filldraw  (-2.5,2) circle (1.5pt) node[left]{\scriptsize $ $};
\filldraw  (-2.5,-2) circle (1.5pt) node[left]{\scriptsize $ $};
\filldraw  (2.5,2) circle (1.5pt) node[left]{\scriptsize $ $};
\filldraw  (2.5,-2) circle (1.5pt) node[left]{\scriptsize $ $};

\filldraw  (-2,2.5) circle (1.5pt) node[left]{\scriptsize $ $};
\filldraw  (-2,-2.5) circle (1.5pt) node[left]{\scriptsize $ $};
\filldraw  (2,2.5) circle (1.5pt) node[left]{\scriptsize $ $};
\filldraw  (2,-2.5) circle (1.5pt) node[left]{\scriptsize $ $};

 \draw[red]  (-2,-1.5)  --  (-2,1.5);
 \draw[red]  (-3,-1.5)  --  (-3,1.5);
 \draw[red]  (2,-1.5)  --  (2,1.5);
 \draw[red]  (3,-1.5)  --  (3,1.5);
 
 \draw[blue]  (-1.5,-2)  --  (1.5,-2);
 \draw[blue]  (-1.5,-3)  --  (1.5,-3);
 \draw[blue]  (-1.5,2)  --  (1.5,2);
 \draw[blue]  (-1.5,3)  --  (1.5,3);

 \draw[red] (-2.5,0) node{$A$};
 \draw[blue] (0,2.5) node{$B$};
 \draw[red] (2.5,0) node{$A$};
 \draw[blue] (0,-2.5) node{$B$};

  \draw[dotted]  (-3,1.5)  --  (-2,1.5);
  \draw[dotted]  (-3,0.9)  --  (-2,0.9);
  \draw[dotted]  (-3,0.3)  --  (-2,0.3);
  \draw[dotted]  (-3,-1.5)  --  (-2,-1.5);
  \draw[dotted]  (-3,-0.9)  --  (-2,-0.9);
  \draw[dotted]  (-3,-0.3)  --  (-2,-0.3);

  \draw[dotted]  (3,1.5)  --  (2,1.5);
  \draw[dotted]  (3,0.9)  --  (2,0.9);
  \draw[dotted]  (3,0.3)  --  (2,0.3);
  \draw[dotted]  (3,-1.5)  --  (2,-1.5);
  \draw[dotted]  (3,-0.9)  --  (2,-0.9);
  \draw[dotted]  (3,-0.3)  --  (2,-0.3);

  \draw[dotted]  (1.5,3)  --  (1.5,2);
  \draw[dotted]  (0.9,3)  --  (0.9,2);
  \draw[dotted]  (0.3,3)  --  (0.3,2);
  \draw[dotted]  (-1.5,3)  --  (-1.5,2);
  \draw[dotted]  (-0.9,3)  --  (-0.9,2);
  \draw[dotted]  (-0.3,3)  --  (-0.3,2);

  \draw[dotted]  (1.5,-3)  --  (1.5,-2);
  \draw[dotted]  (0.9,-3)  --  (0.9,-2);
  \draw[dotted]  (0.3,-3)  --  (0.3,-2);
  \draw[dotted]  (-1.5,-3)  --  (-1.5,-2);
  \draw[dotted]  (-0.9,-3)  --  (-0.9,-2);
  \draw[dotted]  (-0.3,-3)  --  (-0.3,-2);

  \draw[dotted]  (-2,1.5)  --  (-1.5,2);
  \draw[dotted]  (-2,0.9)  --  (-0.9,2);
  \draw[dotted]  (-2,0.3)  --  (-0.3,2);

  \draw[dotted]  (2,1.5)  --  (1.5,2);
  \draw[dotted]  (2,0.9)  --  (0.9,2);
  \draw[dotted]  (2,0.3)  --  (0.3,2);

  \draw[dotted]  (-2,-1.5)  --  (-1.5,-2);
  \draw[dotted]  (-2,-0.9)  --  (-0.9,-2);
  \draw[dotted]  (-2,-0.3)  --  (-0.3,-2);

  \draw[dotted]  (2,-1.5)  --  (1.5,-2);
  \draw[dotted]  (2,-0.9)  --  (0.9,-2);
  \draw[dotted]  (2,-0.3)  --  (0.3,-2);

 \draw (0,1.) node{$\tilde{\rho}_A$};
 \draw (0,-1.) node{$\tilde{\rho}_A$};

 \draw (-3.4,-0.5) node{\scriptsize$0$};
 \draw (-3.4,0.5) node{\scriptsize$2\beta$};
\draw (-1.5,0.3) node{\scriptsize$\beta$};

 \draw (3.4,0.5) node{\scriptsize$2\beta$};
 \draw (3.4,-0.5) node{\scriptsize$4\beta$};
\draw (1.5,0.3) node{\scriptsize$3\beta$};

 \draw (3.8,-2) node{\scriptsize$\chi^A(s)$};
 \draw (2,3.5) node{\scriptsize$\chi^B(s)$};
\end{tikzpicture},\label{contourrhoA2}
\end{equation}
where the symmetry of interchanging $A$ and $B$ becomes obvious. Here we have parametrized the contour by $s \in [0,4\beta]$ anticlockwise. We define $\chi^S(s)\propto\chi^S_2(s)$ for $s\in[\beta/2,3\beta/2]\cup[5\beta/2,7\beta/2]$ and $\chi^S(s)\propto\chi^S_1(s)$ otherwise. Similar to the previous section, the boundary condition again becomes the traditional continuous and anti-periodic boundary condition for $\chi^S$:
\begin{equation}
\begin{aligned}
\chi^A(0^+)&=-\chi^A(2\beta^-),\ \ \ \ \ \ \chi^A(2\beta^+)=-\chi^A(4\beta^-),\\
\chi^B(0^+)&=-\chi^B(4\beta^-),\ \ \ \ \ \ \chi^B(\beta^-)=\chi^B(3\beta^+),
\ \ \ \ \ \ \chi^B(\beta^+)&=-\chi^B(3\beta^-).
\end{aligned}
\end{equation}
Similar to the previous subsection, we consider the Green's function for $\chi^S$: $$G^{A/B}(s,s')=\left<\mathcal{T}_{\mathcal{C}}\chi^{A/B}(s)\chi^{A/B}(s')\right>.$$The Schwinger-Dyson equation then reads
\begin{equation}
\begin{aligned}
G^A(s,s')&=(\partial_\tau-\Sigma^A)^{-1}_A(s,s'),\ \ \ \ \ \ G^B(s,s')=(\partial_\tau-\Sigma^B)^{-1}_B(s,s'),\\
\Sigma^A(s,s')&=\Sigma^B(s,s')=J^2(\lambda G^A(s,s')+(1-\lambda)G^B(s,s'))^{q-1}P^{(2)}(s,s').\label{SDKMe}
\end{aligned}
\end{equation}
Here the $A/B$ labels different boundary conditions for $A/B$ subsystem. This self-energy can be directly understood by the melon diagram (Here $q=4$ for example):
\begin{equation}
\Sigma^A(s,s')=\Sigma^B(s,s')=\begin{tikzpicture}[baseline={([yshift=-6pt]current bounding box.center)}, scale=1.1]
\draw[thick] (-30pt,0pt) -- (-20pt,0pt);
\draw[dashed,thick] (-20pt,0pt)..controls (-16pt,28pt) and (16pt,28pt)..(20pt,0pt);
\draw[thick] (-20pt,0pt)..controls (-16pt,16pt) and (16pt,16pt)..(20pt,0pt);
\draw[thick] (-20pt,0pt)..controls (-16pt,-16pt) and (16pt,-16pt)..(20pt,0pt);
\draw[thick] (-20pt,0pt)--(20pt,0pt);
\draw[thick] (20pt,0pt) -- (30pt,0pt);

\draw (-22pt,-6pt) node{$s$};
\draw (22pt,-6pt) node{$s'$};
\draw (0,15pt) node{\scriptsize$A/B$};
\draw (0,4pt) node{\scriptsize$A/B$};
\draw (0,-16pt) node{\scriptsize$A/B$};
\end{tikzpicture}.
\end{equation}
with $\lambda$ or $1-\lambda$ being the probability of having a mode in subsystem $A$ or $B$. We have $P^{(2)}(s,s')=1$ if both $\chi^S(s)$ and $\chi^S(s')$ represents the same field ($\chi^S_1$ or $\chi^S_2$) and otherwise zero. Note that the main difference between this expression of self-energy and that for computing the subsystem R\'enyi entropy of thermal ensembles \cite{new} is the presence of $P^{(2)}$. 

These set of equations for $G^{A/B}$ and $\Sigma^{A/B}$ can also be directly derived by writing out the $G-\Sigma$ action and taking the saddle point approximation. Consequently, after solving \eqref{SDKMe}, we have $\text{tr}_A \tilde{\rho}_A^2=e^{-I^{(2)}}$ with:
\begin{equation}
\begin{aligned}
\frac{I^{(2)}}{N}=&\frac{\lambda}{4}\log \det G^A(s,s')+\frac{\lambda(q-1)}{4q}\int ds ds'G^A(s,s')\Sigma^A(s,s')+\\&\frac{1-\lambda}{4}\log \det G^B(s,s') 
+\frac{(1-\lambda)(q-1)}{4q}\int ds ds'G^B(s,s')\Sigma^B(s,s'). \label{on shell action KMe}
\end{aligned}
\end{equation} 
We have $I^{(2)}(\lambda=0)=I^{(2)}(\lambda=1)=2I_{\text{KM}}$. The second R\'enyi entanglement entropy is then given by
\begin{equation}
\mathcal{S}_A^{(2)}=-\log(\text{tr}_A\rho_A^2)=-\log\left(\frac{\text{tr}_A\tilde{\rho}_A^2}{Z_{\text{KM}}^2}\right)=I^{(2)}-2I_{\text{KM}}.
\end{equation}
The generalization of above discussions to $n$-th R\'enyi entropy is straightforward.

\section{Numerical results}\label{sec4}
Because of the lack of translational invariance, the analytical study of \eqref{SDKMe} is difficult. In this section, we present numerical results for the entanglement entropy with different $q$ and $T/J$. 

We perform the numerical iteration of \eqref{SDKMe} similar to that in \cite{Yiming,new}. We discretize the time $s$ into $L$ points, with $d s=4\beta/L$. For $\beta =50$, we typically take $L\sim 400\sim 600$. The equation \eqref{SDKMe} then becomes a matrix equation:
\begin{equation}
\begin{aligned}
\left(G^A\right)_{ij}&=\left((G^A_0)^{-1}-\Sigma^A\right)^{-1}_{ij},\ \ \ \ \ \ \left(G^B\right)_{ij}=\left((G^B_0)^{-1}-\Sigma^B\right)^{-1}_{ij},\\
\left(\Sigma^A\right)_{ij}&=\left(\Sigma^B\right)_{ij}=J^2ds^2\left(\lambda \left(G^A\right)_{ij}+(1-\lambda)\left(G^B\right)_{ij}\right)^{q-1}P^{(2)}_{ij}.
\end{aligned}
\end{equation}
Here $G^A_0$ and $G^B_0$ are the Green's functions without interaction $J$ on the contour \eqref{contourrhoA2}. Their elements are either $\pm 1/2$ or $0$ depending on the time ordering and the connectivity of contours. Explicitly:
\begin{equation}
\begin{aligned}
\left(G^A_0\right)_{ij}&=\frac{1}{2}\text{sign}(i-j)\ \ \ \ \ \ \text{for}\ \{i,j\}\subset [1,L/2]\ \text{or}\ [L/2+1,L],\\
\left(G^B_0\right)_{ij}&=\frac{1}{2}\text{sign}(i-j)\ \ \ \ \ \ \text{for}\ \{i,j\}\subset [1,L/4]\cup[3L/4+1,L]  \ \text{or}\ [L/4+1,3L/4].
\end{aligned}
\end{equation}
The on-shell action is then 
\begin{equation}
\begin{aligned}
\frac{I^{(2)}}{N}=&\frac{\lambda}{4}\log \det \left[G^A(G^A_0)^{-1}\right]+\frac{\lambda(q-1)}{4q}\text{tr} \left[G^A(\Sigma^A)^T\right]-\frac{1}{2}\log 2\\
&+\frac{1-\lambda}{4}\log \det \left[G^B(G^B_0)^{-1}\right]
+\frac{(1-\lambda)(q-1)}{4q}\text{tr} \left[G^B(\Sigma^B)^T\right]. \label{on shell action KMenum}
\end{aligned}
\end{equation} 
Here we have used $\log\det(G^A_0)^{-1}=\log\det(G^B_0)^{-1}=2\log 2$ to enable the convergence. The extrapolation towards $1/L\rightarrow 0$ is performed finally.

\begin{figure}[t]
  \center
  \includegraphics[width=1\columnwidth]{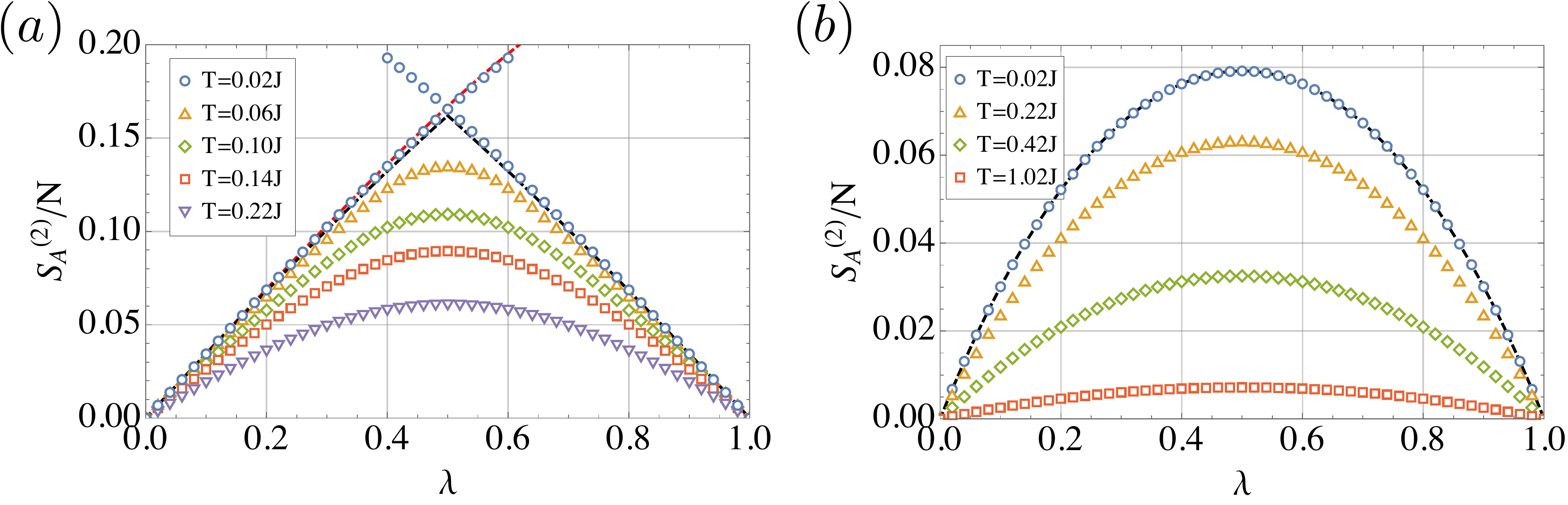}
  \caption{(a). The entanglement entropy $\mathcal{S}_A^{(2)}/N$ of KM states at different temperature $T/J$ with $q=4$. The black dashed line is the analytical approximation for \eqref{ETH}: $\mathcal{S}_A^{(2)}(\lambda)/N=x (\log(2)/2 -  \arcsin (x^{3/2})/16 )$ with $x=\text{min}\{\lambda,1-\lambda\}$. The red dashed line is the subsystem entropy for a thermal ensemble with $\beta J=50$ \cite{new}. (b). The entanglement entropy $\mathcal{S}_A^{(2)}/N$ of KM states at different temperature $T/J$ with $q=2$. The black dashed line is the analytical formula for the SYK$_2$ ground state \cite{new}.}\label{fig1}
 \end{figure}

Now we present numerical results for the entanglement entropy of KM pure states. We first focus on the SYK$_q$ model with $q=4$ or $q=2$ as an example of strongly interacting systems or non-interacting systems.

The result of $\mathcal{S}_A^{(2)}$ for different $\beta J$ with $q=4$ is shown in Figure \ref{fig1} (a). We have also plotted the analytical approximation \cite{huang2019eigenstate} as the black dashed line and the subsystem entropy for a thermal ensemble with $\beta J=50$ \cite{new} as the red dashed line. For small $\beta J$, the entanglement builds up quickly as $\beta J$ increases, and $\mathcal{S}_A^{(2)}/N$ is an analytical function of $\lambda$. On the other side, for large $\beta J \gtrsim 30$ near $\lambda \sim 1/2$, there exist two different saddle point solutions, and the coexistence region becomes larger as $\beta J$ increases. The true curve for $\mathcal{S}_A^{(2)}$ is determined by the comparing actions of different saddles, leading a first-order transition. As we will see in the next section, these two saddle points smoothly connected to the subsystem entropy $\mathcal{S}_{A,\text{th}}^{(2)}$ or $\mathcal{S}_{B,\text{th}}^{(2)}$ of a thermal ensemble at corresponding temperature. For $T/J=0.02$, the numerical result of the KM state gives $\mathcal{S}_A^{(2)}(1/2)=0.331M$, which is larger than the analytical approximation $\mathcal{S}_A^{(2)}(1/2)=0.324M$, but also a little smaller than the thermal ensemble result $\mathcal{S}_A^{(2)}(1/2)=0.334M$. We attribute this to the fact that KM states are non-thermal. This is to be compared with the quadratic Hamiltonian case with $q=2$ shown in Figure \ref{fig1} (b), where $\mathcal{S}_A^{(2)}(\lambda)$ is analytical even for very low temperature.

The existence of the transition gives rise to the singularity of the Page curve at $\lambda=1/2$ \cite{page1993average}, as expected for general chaotic systems. Consequently, the existence of transition in highly entangled pure states to be a general feature for interacting systems with saddle-point description, including different generalizations of the SYK model.

\begin{figure}[t]
  \center
  \includegraphics[width=1\columnwidth]{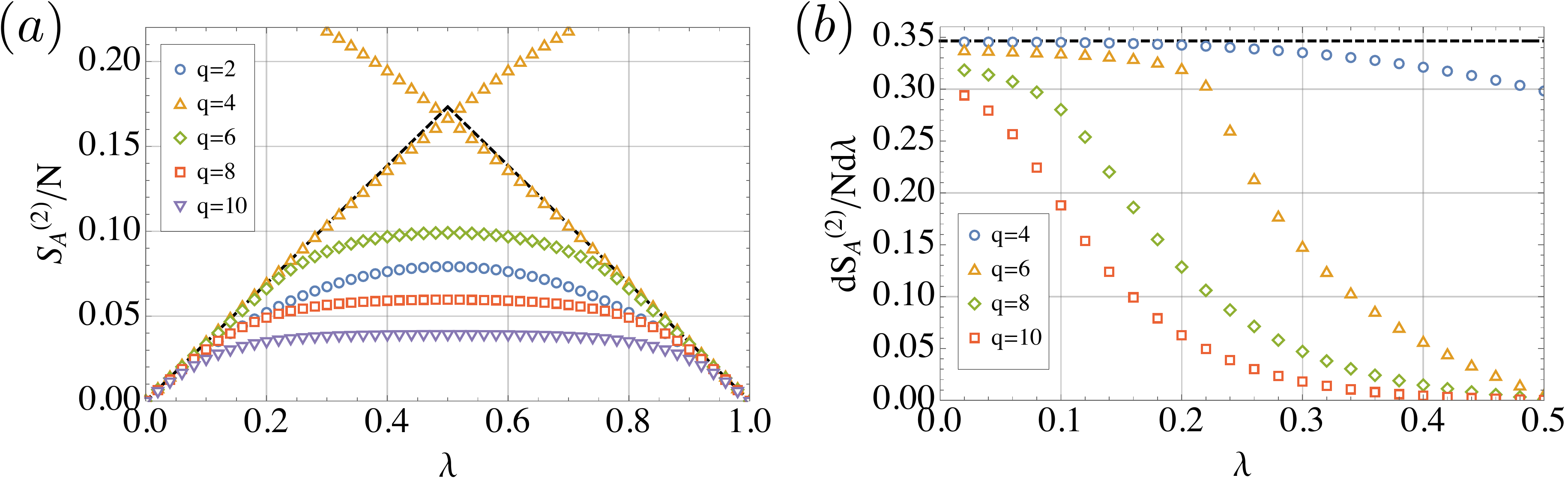}
  \caption{(a). The entanglement entropy $\mathcal{S}_A^{(2)}/N$ of KM states at different $q$ with $\beta \mathcal{J}=50$. The black dashed line is the maximal entropy: $\mathcal{S}_A^{(2)}(\lambda)/N=x \log(2)/2 $ with $x=\text{min}\{\lambda,1-\lambda\}$. (b). The derivative of the entanglement entropy $d\mathcal{S}_A^{(2)}/Nd\lambda$ of KM states at different $q$ with $\beta \mathcal{J}=50$. The black dashed line represent the maximal entropy $\log (2)/2$. }\label{fig2}
 \end{figure}

We then consider $q$ dependence for $\mathcal{S}_A^{(2)}$. Here we fix $\beta\mathcal{J}=50$. The numerical results are shown in Figure \ref{fig2} (a). The $q=4$ and $q=2$ case has been discussed above. If we further increase $q\geq 6$, we find the entanglement entropy decreases rapidly as $q$ increases. This can be understood that in the large $q$ limit, the system is weakly interacting. As a result, for fixed $\beta \mathcal{J}$ the system becomes less entangled as $q$ increases. It is also interesting to notice that for larger $q$, $\mathcal{S}_A^{(2)}$ becomes much more flat near $\lambda \sim 1/2$, which can also be seen from Figure \ref{fig2} (b). There is an analogy phenomenon for the entanglement entropy under the random Hamiltonian evolution \cite{you2018entanglement}. It is also reasonable that for larger $q$, the Hamiltonian becomes denser and resembles a random Hamiltonian.

\section{Quench dynamics for pure-state entanglement entropy}\label{sec5}
We now turn to the study of real-time entanglement dynamics by considering the evolution of $|\text{KM}\rangle$. The normalized state after evolving time $t$ is 
\begin{equation}
|\psi(t)\rangle=
\frac{1}{\sqrt{Z_{\text{KM}}}}e^{-iHt}|\text{KM}\rangle.
\end{equation}
We first consider the two-point function of $\chi_i$ on real-time. There are studies on quench dynamics for two-point functions for different SYK-like models in the large-$N$ limit \cite{eberlein2017quantum,haldar2019quench,kuhlenkamp2019periodically,zhang2019evaporation,almheiri2019universal} by solving the Kadanoff-Baym equation \cite{stefanucci2013nonequilibrium} \footnote{In these works, the initial state is a thermal ensemble. Here we instead of focus on pure states.}. However, in our case the two point function can be directly obtained by  an analytical continuation of \eqref{Gij}. Consequently, the diagonal components $G_{ii}$ are again thermal at any time while the off-diagonal components $G_{2j-1,2j}(t)\sim G_{\text{th}}(\beta/2)^2e^{-2v t}$ in the long-time limit. Here $v$ is the decay rate of the real-time two-point function on thermal ensemble. In the low-temperature limit, we have $v\sim 1/\beta$ and $G_{\text{th}}(\beta/2) \rightarrow 0$. In the high-temperature limit, we instead have $v\sim J$, which leads to a thermalization time $t_{\text{th}}\sim 1/J$.

We would like to study this quenching process from the entanglement perspective. To compute the evolution of the entanglement entropy, we again apply the path integral representation using the contour in \eqref{contourrhoA2}. The main difference is that now the solid lines can be either (forward/backward) real or imaginary-time evolution:
\begin{equation}
\text{tr}_A \rho_A(t)^2=\text{tr}_A (\text{tr}_B |\psi(t) \rangle \langle \psi(t)|)^2=\frac{1}{Z_{\text{KM}}^2}\times\begin{tikzpicture}[thick,scale = 0.45,baseline={([yshift=-4pt]current bounding box.center)}]
 \draw[red] (-2,-1.5) arc(0:-180:0.5 and 0.5);
 \draw[red] (-2,1.5) arc(0:180:0.5 and 0.5);
 \draw[red] (2,-1.5) arc(-180:0:0.5 and 0.5);
 \draw[red] (2,1.5) arc(180:0:0.5 and 0.5);
 \draw[blue] (-1.5,2) arc(-90:-270:0.5 and 0.5);
 \draw[blue] (-1.5,-2) arc(90:270:0.5 and 0.5);
 \draw[blue] (1.5,2) arc(-90:90:0.5 and 0.5);
 \draw[blue] (1.5,-2) arc(90:-90:0.5 and 0.5);

\filldraw  (-2.5,2) circle (1.5pt) node[left]{\scriptsize $ $};
\filldraw  (-2.5,-2) circle (1.5pt) node[left]{\scriptsize $ $};
\filldraw  (2.5,2) circle (1.5pt) node[left]{\scriptsize $ $};
\filldraw  (2.5,-2) circle (1.5pt) node[left]{\scriptsize $ $};

\filldraw  (-2,2.5) circle (1.5pt) node[left]{\scriptsize $ $};
\filldraw  (-2,-2.5) circle (1.5pt) node[left]{\scriptsize $ $};
\filldraw  (2,2.5) circle (1.5pt) node[left]{\scriptsize $ $};
\filldraw  (2,-2.5) circle (1.5pt) node[left]{\scriptsize $ $};

 \draw[red]  (-2,-1.5)  --  (-2,1.5);
 \draw[red]  (-3,-1.5)  --  (-3,1.5);
 \draw[red]  (2,-1.5)  --  (2,1.5);
 \draw[red]  (3,-1.5)  --  (3,1.5);
 
 \draw[blue]  (-1.5,-2)  --  (1.5,-2);
 \draw[blue]  (-1.5,-3)  --  (1.5,-3);
 \draw[blue]  (-1.5,2)  --  (1.5,2);
 \draw[blue]  (-1.5,3)  --  (1.5,3);

 \draw[red] (-2.5,0) node{$A$};
 \draw[blue] (0,2.5) node{$B$};
 \draw[red] (2.5,0) node{$A$};
 \draw[blue] (0,-2.5) node{$B$};

  \draw[dotted]  (-3,1.5)  --  (-2,1.5);
  \draw[dotted]  (-3,0.9)  --  (-2,0.9);
  \draw[dotted]  (-3,0.3)  --  (-2,0.3);
  \draw[dotted]  (-3,-1.5)  --  (-2,-1.5);
  \draw[dotted]  (-3,-0.9)  --  (-2,-0.9);
  \draw[dotted]  (-3,-0.3)  --  (-2,-0.3);

  \draw[dotted]  (3,1.5)  --  (2,1.5);
  \draw[dotted]  (3,0.9)  --  (2,0.9);
  \draw[dotted]  (3,0.3)  --  (2,0.3);
  \draw[dotted]  (3,-1.5)  --  (2,-1.5);
  \draw[dotted]  (3,-0.9)  --  (2,-0.9);
  \draw[dotted]  (3,-0.3)  --  (2,-0.3);

  \draw[dotted]  (1.5,3)  --  (1.5,2);
  \draw[dotted]  (0.9,3)  --  (0.9,2);
  \draw[dotted]  (0.3,3)  --  (0.3,2);
  \draw[dotted]  (-1.5,3)  --  (-1.5,2);
  \draw[dotted]  (-0.9,3)  --  (-0.9,2);
  \draw[dotted]  (-0.3,3)  --  (-0.3,2);

  \draw[dotted]  (1.5,-3)  --  (1.5,-2);
  \draw[dotted]  (0.9,-3)  --  (0.9,-2);
  \draw[dotted]  (0.3,-3)  --  (0.3,-2);
  \draw[dotted]  (-1.5,-3)  --  (-1.5,-2);
  \draw[dotted]  (-0.9,-3)  --  (-0.9,-2);
  \draw[dotted]  (-0.3,-3)  --  (-0.3,-2);

  \draw[dotted]  (-2,1.5)  --  (-1.5,2);
  \draw[dotted]  (-2,0.9)  --  (-0.9,2);
  \draw[dotted]  (-2,0.3)  --  (-0.3,2);

  \draw[dotted]  (2,1.5)  --  (1.5,2);
  \draw[dotted]  (2,0.9)  --  (0.9,2);
  \draw[dotted]  (2,0.3)  --  (0.3,2);

  \draw[dotted]  (-2,-1.5)  --  (-1.5,-2);
  \draw[dotted]  (-2,-0.9)  --  (-0.9,-2);
  \draw[dotted]  (-2,-0.3)  --  (-0.3,-2);

  \draw[dotted]  (2,-1.5)  --  (1.5,-2);
  \draw[dotted]  (2,-0.9)  --  (0.9,-2);
  \draw[dotted]  (2,-0.3)  --  (0.3,-2);

  \draw[thick]  (-2.1,0)  --  (-1.9,0);
  \draw[thick]  (-3.1,0)  --  (-2.9,0);
  \draw[thick]  (2.1,0)  --  (1.9,0);
  \draw[thick]  (3.1,0)  --  (2.9,0);

  \draw[thick]  (-2.1,0.9)  --  (-1.9,0.9);
  \draw[thick]  (-3.1,0.9)  --  (-2.9,0.9);
  \draw[thick]  (2.1,0.9)  --  (1.9,0.9);
  \draw[thick]  (3.1,0.9)  --  (2.9,0.9);

  \draw[thick]  (-2.1,-0.9)  --  (-1.9,-0.9);
  \draw[thick]  (-3.1,-0.9)  --  (-2.9,-0.9);
  \draw[thick]  (2.1,-0.9)  --  (1.9,-0.9);
  \draw[thick]  (3.1,-0.9)  --  (2.9,-0.9);

  \draw[thick]  (0,-2.1)  --  (0,-1.9);
  \draw[thick]  (0,-3.1)  --  (0,-2.9);
  \draw[thick]  (0,2.1)  --  (0,1.9);
  \draw[thick]  (0,3.1)  --  (0,2.9);

  \draw[thick]  (0.9,-2.1)  --  (0.9,-1.9);
  \draw[thick]  (0.9,-3.1)  --  (0.9,-2.9);
  \draw[thick]  (0.9,2.1)  --  (0.9,1.9);
  \draw[thick]  (0.9,3.1)  --  (0.9,2.9);

  \draw[thick]  (-0.9,-2.1)  --  (-0.9,-1.9);
  \draw[thick]  (-0.9,-3.1)  --  (-0.9,-2.9);
  \draw[thick]  (-0.9,2.1)  --  (-0.9,1.9);
  \draw[thick]  (-0.9,3.1)  --  (-0.9,2.9);

  \draw[mid arrow, red] (-3,-0.8) -- (-3,0);
  \draw[mid arrow, red] (-2,-0.8) -- (-2,0);
  \draw[mid arrow, red] (3,-0.8) -- (3,0);
  \draw[mid arrow, red] (2,-0.8) -- (2,0);

  \draw[mid arrow, red] (-3,0.8) -- (-3,0);
  \draw[mid arrow, red] (-2,0.8) -- (-2,0);
  \draw[mid arrow, red] (3,0.8) -- (3,0);
  \draw[mid arrow, red] (2,0.8) -- (2,0);

  \draw[mid arrow, blue] (-0.8,-3) -- (0,-3);
  \draw[mid arrow, blue] (-0.8,-2) -- (0,-2);
  \draw[mid arrow, blue] (-0.8,3) -- (0,3);
  \draw[mid arrow, blue] (-0.8,2) -- (0,2);

  \draw[mid arrow, blue] (0.8,-3) -- (0,-3);
  \draw[mid arrow, blue] (0.8,-2) -- (0,-2);
  \draw[mid arrow, blue] (0.8,3) -- (0,3);
  \draw[mid arrow, blue] (0.8,2) -- (0,2);

\end{tikzpicture},\label{contourrhoA2t}
\end{equation}
Here we draw an arrow for the direction of real-time evolutions. When the parametrization is along the same direction as the arrow, the real-time evolution is effectively forward. Otherwise, the evolution is backward. Since the evolution is governed by $e^{-\tau H}$ for imaginary-time evolution and $e^{\mp i H t}$ for forward/backward real-time evolution, we need to add additional factor of $\pm i$. This leads to the modification of \eqref{SDKMe}:
\begin{equation}
\begin{aligned}
G^A(s,s')&=(\partial_\tau-\Sigma^A)^{-1}_A(s,s'),\ \ \ \ \ \ G^B(s,s')=(\partial_\tau-\Sigma^B)^{-1}_B(s,s'),\\
\Sigma^A(s,s')&=\Sigma^B(s,s')=J^2(\lambda G^A(s,s')+(1-\lambda)G^B(s,s'))^{q-1}\tilde{P}^{(2)}(s,s').\label{SDKMt}
\end{aligned}
\end{equation}
where now we have $s \in [0,4(\beta+2t))$, including both the real-time and the imaginary-time evolution. Here $\tilde{P}^{(2)}(s,s')=P^{(2)}(s,s') f(s)f(s')$ with $f(s)=1,i$ or $-i$ for $s$ being a parameter of an imaginary-/forward real- or backward real-time evolution. After solving \eqref{SDKMt}, the on-shell action can still be computed as \eqref{on shell action KMe}. 

We first consider the case with $q\geq4$. Since the system satisfy the ETH, we expect when $\lambda<1/2$ the entanglement entropy approaches the subsystem entropy of a thermal ensemble in the long time limit. This can be understood by expanding $$|\psi(t)\rangle=\sum_mc_me^{-iE_mt}|E_m\rangle.$$ Here $|E_m\rangle$ are eigenstates of the Hamiltonian with eigenenergy $E_m$. The entanglement entropy for a subsystem $A$ can be written as the expectation of the swap operator on two replicas of the original system:
\begin{equation}
e^{-\mathcal{S}_A^{(2)}(t)}=\text{tr}_A (\text{tr}_B |\psi(t) \rangle \langle \psi(t)|)^2=\text{tr}\left[\hat{S}^A|\psi(t) \rangle \langle \psi(t)| \otimes |\psi(t) \rangle \langle \psi(t)|\right]
\end{equation}
Here $\hat{S}^A$ is the swap operator of subsystem $A$ on doubled Hilbert space. Consequently, in the long time limit, we have
\begin{equation}
\begin{aligned}
e^{-\mathcal{S}_A^{(2)}}\approx&\sum_{m,n}|c_mc_n|^2\text{tr}\left[\hat{S}^A\left(|E_m \rangle \langle E_m| \otimes |E_n \rangle \langle E_n|+|E_n \rangle \langle E_m| \otimes |E_m \rangle \langle E_n|\right) \right]\\
=&\sum_{m,n}|c_mc_n|^2\text{tr}\left[(\hat{S}^A+\hat{S}^B)|E_m \rangle \langle E_m| \otimes |E_n \rangle \langle E_n|\right]\\
=&\text{tr}\left[(\hat{S}^A+\hat{S}^B)\rho_{\text{th}} \otimes\rho_{\text{th}}\right]=e^{-\mathcal{S}_{A,th}^{(2)}}+e^{-\mathcal{S}_{B,th}^{(2)}}.\label{dynamics}
\end{aligned}
\end{equation}
We have used the fact that the diagonal ensemble $\sum_n|c_n|^2|E_n\rangle \langle E_n|$ is approximately a thermal ensemble $\rho_{\text{th}}$ (of the full system) with inverse temperature $\beta$ for systems satisfying the ETH. Previous study shows $\mathcal{S}_{A,th}^{(2)}<\mathcal{S}_{B,th}^{(2)}$ for $\lambda<1/2$ while $\mathcal{S}_{A,th}^{(2)}>\mathcal{S}_{B,th}^{(2)}$ for $\lambda>1/2$ \cite{new}. Note that this is a non-trival statement since the effective temperature depends on $\lambda$ for non-local Hamiltonians \eqref{ETH}. Consequently, in the large-$N$ limit, we have $\mathcal{S}_A^{(2)}=\mathcal{S}_{A,th}^{(2)}$ for $\lambda<1/2$ and $\mathcal{S}_A^{(2)}=\mathcal{S}_{B,th}^{(2)}$ for $\lambda>1/2$.

\begin{figure}[t]
  \center
  \includegraphics[width=1\columnwidth]{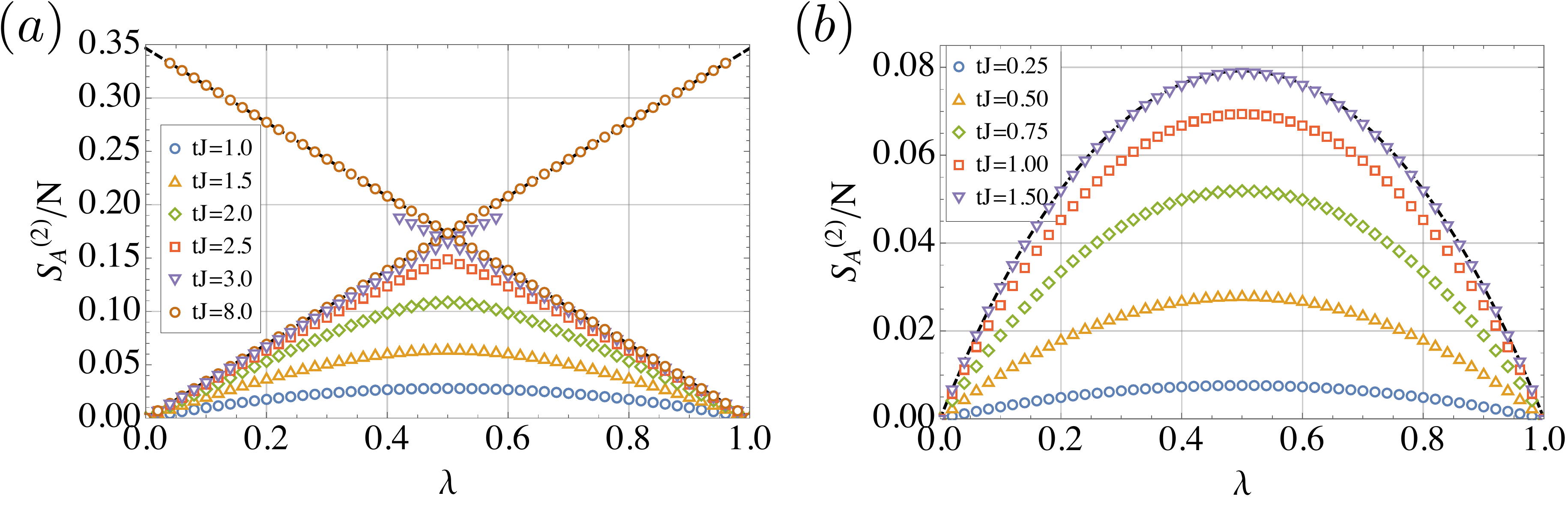}
  \caption{(a). The dynamics of the R\'enyi entanglement entropy $\mathcal{S}_A^{(2)}(t)$ as a function of subsystem size $\lambda$ with $q=4$ and $\beta=0$. The black dashed reference line is $\lambda \log(2)/2$ and $(1-\lambda)\log(2)/2$. (b). The dynamics of the R\'enyi entanglement entropy $\mathcal{S}_A^{(2)}(t)$ as a function of subsystem size $\lambda$ with $q=2$ and $\beta=0$. The black dashed line is the analytical formula for the SYK$_2$ ground state \cite{new}. }\label{fig3}
 \end{figure}

The numerical result for $\mathcal{S}_{A}^{(2)}(t)$ with $q=4$ is shown in Figure \ref{fig3} (a). Here we have set $\beta=0$. Similar to tunning the imaginary-time $\beta$, we find a first-order transition in the long real-time limit. By comparing with $\lambda \log(2)/2$ and $(1-\lambda)\log(2)/2$, two different saddles in the long-time limit can be identified with the contribution from $e^{-\mathcal{S}_{A,th}^{(2)}}$ or $e^{-\mathcal{S}_{B,th}^{(2)}}$ in \eqref{dynamics}. We have also checked the match for finite $\beta$.

We could also understand the two saddle points directly from the path-integral representation \eqref{contourrhoA2t}. For convenience, here we put back the possible $\{s\}$ dependence of KM states. we write $|\{s\}\rangle=|\{s_A\}\rangle_A \otimes |\{s_B\}\rangle_B$. In the long-time limit, the saddle point solutions can be understood as follows: We first consider turn off the interaction between subsystem $A$ and $B$. Then the Green's functions $G^A$ and $G^B$ become block diagonal according to $G^A_0/G^B_0$. When we turn on the interaction between subsystem $A$ and $B$, one saddle is given by: We keep $G^B$ almost replica diagonal. Two half contours of $G^A$ that interacts with the same $B$ subsystem become effectively connected \cite{saad2018semiclassical,penington2019replica,Yiming}, while two $\chi^A_i$ interact with different $B$ subsystems becomes less correlated far away from the boundary of the contour. Graphically, this means
\begin{equation}
\text{Saddle point corresponds to $\mathcal{S}_{A,th}^{(2)}$:}\ \ \ \ \ \ G^A(s_1,s_2)=\begin{tikzpicture}[thick,scale = 0.45,baseline={([yshift=-4pt]current bounding box.center)}]
 \draw[red] (-2,-1.5) arc(0:-180:0.5 and 0.5);
 \draw[red] (-2,1.5) arc(0:180:0.5 and 0.5);
 \draw[red] (2,-1.5) arc(-180:0:0.5 and 0.5);
 \draw[red] (2,1.5) arc(180:0:0.5 and 0.5);
 \draw[blue] (-1.5,2) arc(-90:-270:0.5 and 0.5);
 \draw[blue] (-1.5,-2) arc(90:270:0.5 and 0.5);
 \draw[blue] (1.5,2) arc(-90:90:0.5 and 0.5);
 \draw[blue] (1.5,-2) arc(90:-90:0.5 and 0.5);

\filldraw  (-2.5,2) circle (1.5pt) node[left]{\scriptsize $ $};
\filldraw  (-2.5,-2) circle (1.5pt) node[left]{\scriptsize $ $};
\filldraw  (2.5,2) circle (1.5pt) node[left]{\scriptsize $ $};
\filldraw  (2.5,-2) circle (1.5pt) node[left]{\scriptsize $ $};

\filldraw  (-2,2.5) circle (1.5pt) node[left]{\scriptsize $ $};
\filldraw  (-2,-2.5) circle (1.5pt) node[left]{\scriptsize $ $};
\filldraw  (2,2.5) circle (1.5pt) node[left]{\scriptsize $ $};
\filldraw  (2,-2.5) circle (1.5pt) node[left]{\scriptsize $ $};

 \draw[red]  (-2,-1.5)  --  (-2,1.5);
 \draw[red]  (-3,-1.5)  --  (-3,1.5);
 \draw[red]  (2,-1.5)  --  (2,1.5);
 \draw[red]  (3,-1.5)  --  (3,1.5);
 
 \draw[blue]  (-1.5,-2)  --  (1.5,-2);
 \draw[blue]  (-1.5,-3)  --  (1.5,-3);
 \draw[blue]  (-1.5,2)  --  (1.5,2);
 \draw[blue]  (-1.5,3)  --  (1.5,3);

 \draw[red] (-2.5,0) node{$A$};
 \draw[blue] (0,2.5) node{$B$};
 \draw[red] (2.5,0) node{$A$};
 \draw[blue] (0,-2.5) node{$B$};

  \draw[dotted]  (-3,1.5)  --  (-2,1.5);
  \draw[dotted]  (-3,0.9)  --  (-2,0.9);
  \draw[dotted]  (-3,0.3)  --  (-2,0.3);
  \draw[dotted]  (-3,-1.5)  --  (-2,-1.5);
  \draw[dotted]  (-3,-0.9)  --  (-2,-0.9);
  \draw[dotted]  (-3,-0.3)  --  (-2,-0.3);

  \draw[dotted]  (3,1.5)  --  (2,1.5);
  \draw[dotted]  (3,0.9)  --  (2,0.9);
  \draw[dotted]  (3,0.3)  --  (2,0.3);
  \draw[dotted]  (3,-1.5)  --  (2,-1.5);
  \draw[dotted]  (3,-0.9)  --  (2,-0.9);
  \draw[dotted]  (3,-0.3)  --  (2,-0.3);

  \draw[dotted]  (1.5,3)  --  (1.5,2);
  \draw[dotted]  (0.9,3)  --  (0.9,2);
  \draw[dotted]  (0.3,3)  --  (0.3,2);
  \draw[dotted]  (-1.5,3)  --  (-1.5,2);
  \draw[dotted]  (-0.9,3)  --  (-0.9,2);
  \draw[dotted]  (-0.3,3)  --  (-0.3,2);

  \draw[dotted]  (1.5,-3)  --  (1.5,-2);
  \draw[dotted]  (0.9,-3)  --  (0.9,-2);
  \draw[dotted]  (0.3,-3)  --  (0.3,-2);
  \draw[dotted]  (-1.5,-3)  --  (-1.5,-2);
  \draw[dotted]  (-0.9,-3)  --  (-0.9,-2);
  \draw[dotted]  (-0.3,-3)  --  (-0.3,-2);

  \draw[dotted]  (-2,1.5)  --  (-1.5,2);
  \draw[dotted]  (-2,0.9)  --  (-0.9,2);
  \draw[dotted]  (-2,0.3)  --  (-0.3,2);

  \draw[dotted]  (2,1.5)  --  (1.5,2);
  \draw[dotted]  (2,0.9)  --  (0.9,2);
  \draw[dotted]  (2,0.3)  --  (0.3,2);

  \draw[dotted]  (-2,-1.5)  --  (-1.5,-2);
  \draw[dotted]  (-2,-0.9)  --  (-0.9,-2);
  \draw[dotted]  (-2,-0.3)  --  (-0.3,-2);

  \draw[dotted]  (2,-1.5)  --  (1.5,-2);
  \draw[dotted]  (2,-0.9)  --  (0.9,-2);
  \draw[dotted]  (2,-0.3)  --  (0.3,-2);

  \draw[thick]  (-2.1,0)  --  (-1.9,0);
  \draw[thick]  (-3.1,0)  --  (-2.9,0);
  \draw[thick]  (2.1,0)  --  (1.9,0);
  \draw[thick]  (3.1,0)  --  (2.9,0);

  \draw[thick]  (-2.1,0.9)  --  (-1.9,0.9);
  \draw[thick]  (-3.1,0.9)  --  (-2.9,0.9);
  \draw[thick]  (2.1,0.9)  --  (1.9,0.9);
  \draw[thick]  (3.1,0.9)  --  (2.9,0.9);

  \draw[thick]  (-2.1,-0.9)  --  (-1.9,-0.9);
  \draw[thick]  (-3.1,-0.9)  --  (-2.9,-0.9);
  \draw[thick]  (2.1,-0.9)  --  (1.9,-0.9);
  \draw[thick]  (3.1,-0.9)  --  (2.9,-0.9);

  \draw[thick]  (0,-2.1)  --  (0,-1.9);
  \draw[thick]  (0,-3.1)  --  (0,-2.9);
  \draw[thick]  (0,2.1)  --  (0,1.9);
  \draw[thick]  (0,3.1)  --  (0,2.9);

  \draw[thick]  (0.9,-2.1)  --  (0.9,-1.9);
  \draw[thick]  (0.9,-3.1)  --  (0.9,-2.9);
  \draw[thick]  (0.9,2.1)  --  (0.9,1.9);
  \draw[thick]  (0.9,3.1)  --  (0.9,2.9);

  \draw[thick]  (-0.9,-2.1)  --  (-0.9,-1.9);
  \draw[thick]  (-0.9,-3.1)  --  (-0.9,-2.9);
  \draw[thick]  (-0.9,2.1)  --  (-0.9,1.9);
  \draw[thick]  (-0.9,3.1)  --  (-0.9,2.9);

  \draw[mid arrow, red] (-3,-0.8) -- (-3,0);
  \draw[mid arrow, red] (-2,-0.8) -- (-2,0);
  \draw[mid arrow, red] (3,-0.8) -- (3,0);
  \draw[mid arrow, red] (2,-0.8) -- (2,0);

  \draw[mid arrow, red] (-3,0.8) -- (-3,0);
  \draw[mid arrow, red] (-2,0.8) -- (-2,0);
  \draw[mid arrow, red] (3,0.8) -- (3,0);
  \draw[mid arrow, red] (2,0.8) -- (2,0);

  \draw[mid arrow, blue] (-0.8,-3) -- (0,-3);
  \draw[mid arrow, blue] (-0.8,-2) -- (0,-2);
  \draw[mid arrow, blue] (-0.8,3) -- (0,3);
  \draw[mid arrow, blue] (-0.8,2) -- (0,2);

  \draw[mid arrow, blue] (0.8,-3) -- (0,-3);
  \draw[mid arrow, blue] (0.8,-2) -- (0,-2);
  \draw[mid arrow, blue] (0.8,3) -- (0,3);
  \draw[mid arrow, blue] (0.8,2) -- (0,2);

  \draw[dashed] (-3.8,0) -- (3.8,0);
  \draw[dashed] (-3.8,3.5) -- (3.8,3.5);
  \draw[dashed] (-3.8,-3.5) -- (-3.8,3.5);
  \draw[dashed] (3.8,-3.5) -- (3.8,3.5);
  \draw[dashed] (-3.8,-3.5) -- (3.8,-3.5);

  \filldraw[red]  (-3,-1.5) circle (3pt) node[left]{\scriptsize $ $};
  \filldraw[red]  (-2.8,1.875) circle (3pt) node[left]{\scriptsize $ $};

  \draw[red] (-2.9,2.6) node{\scriptsize $\chi(s_1)$};
    \draw[red] (-2.9,-2.6) node{\scriptsize $\chi(s_2)$};
\end{tikzpicture}\approx 0.
\end{equation}
Here the contours within each box are effectively connected. The red dots represent the insertion of field $\chi^A$. An important observation is that since the correlation is small between two boxes near the black dots where the boundary condition is imposed, we could make the approximation:
\begin{equation}
\begin{tikzpicture}[thick,scale = 0.45,baseline={([yshift=-4pt]current bounding box.center)}]
 \draw[red] (-2,-1.5) arc(0:-180:0.5 and 0.5);
 \draw[red] (-2,1.5) arc(0:180:0.5 and 0.5);
 \draw[red] (2,-1.5) arc(-180:0:0.5 and 0.5);
 \draw[red] (2,1.5) arc(180:0:0.5 and 0.5);
 \draw[blue] (-1.5,2) arc(-90:-270:0.5 and 0.5);
 \draw[blue] (-1.5,-2) arc(90:270:0.5 and 0.5);
 \draw[blue] (1.5,2) arc(-90:90:0.5 and 0.5);
 \draw[blue] (1.5,-2) arc(90:-90:0.5 and 0.5);

\filldraw  (-2.5,2) circle (1.5pt) node[left]{\scriptsize $ $};
\filldraw  (-2.5,-2) circle (1.5pt) node[left]{\scriptsize $ $};
\filldraw  (2.5,2) circle (1.5pt) node[left]{\scriptsize $ $};
\filldraw  (2.5,-2) circle (1.5pt) node[left]{\scriptsize $ $};

\filldraw  (-2,2.5) circle (1.5pt) node[left]{\scriptsize $ $};
\filldraw  (-2,-2.5) circle (1.5pt) node[left]{\scriptsize $ $};
\filldraw  (2,2.5) circle (1.5pt) node[left]{\scriptsize $ $};
\filldraw  (2,-2.5) circle (1.5pt) node[left]{\scriptsize $ $};

 \draw[red]  (-2,-1.5)  --  (-2,1.5);
 \draw[red]  (-3,-1.5)  --  (-3,1.5);
 \draw[red]  (2,-1.5)  --  (2,1.5);
 \draw[red]  (3,-1.5)  --  (3,1.5);
 
 \draw[blue]  (-1.5,-2)  --  (1.5,-2);
 \draw[blue]  (-1.5,-3)  --  (1.5,-3);
 \draw[blue]  (-1.5,2)  --  (1.5,2);
 \draw[blue]  (-1.5,3)  --  (1.5,3);

 \draw[red] (-2.5,0) node{$A$};
 \draw[blue] (0,2.5) node{$B$};
 \draw[red] (2.5,0) node{$A$};
 \draw[blue] (0,-2.5) node{$B$};

  \draw[dotted]  (-3,1.5)  --  (-2,1.5);
  \draw[dotted]  (-3,0.9)  --  (-2,0.9);
  \draw[dotted]  (-3,0.3)  --  (-2,0.3);
  \draw[dotted]  (-3,-1.5)  --  (-2,-1.5);
  \draw[dotted]  (-3,-0.9)  --  (-2,-0.9);
  \draw[dotted]  (-3,-0.3)  --  (-2,-0.3);

  \draw[dotted]  (3,1.5)  --  (2,1.5);
  \draw[dotted]  (3,0.9)  --  (2,0.9);
  \draw[dotted]  (3,0.3)  --  (2,0.3);
  \draw[dotted]  (3,-1.5)  --  (2,-1.5);
  \draw[dotted]  (3,-0.9)  --  (2,-0.9);
  \draw[dotted]  (3,-0.3)  --  (2,-0.3);

  \draw[dotted]  (1.5,3)  --  (1.5,2);
  \draw[dotted]  (0.9,3)  --  (0.9,2);
  \draw[dotted]  (0.3,3)  --  (0.3,2);
  \draw[dotted]  (-1.5,3)  --  (-1.5,2);
  \draw[dotted]  (-0.9,3)  --  (-0.9,2);
  \draw[dotted]  (-0.3,3)  --  (-0.3,2);

  \draw[dotted]  (1.5,-3)  --  (1.5,-2);
  \draw[dotted]  (0.9,-3)  --  (0.9,-2);
  \draw[dotted]  (0.3,-3)  --  (0.3,-2);
  \draw[dotted]  (-1.5,-3)  --  (-1.5,-2);
  \draw[dotted]  (-0.9,-3)  --  (-0.9,-2);
  \draw[dotted]  (-0.3,-3)  --  (-0.3,-2);

  \draw[dotted]  (-2,1.5)  --  (-1.5,2);
  \draw[dotted]  (-2,0.9)  --  (-0.9,2);
  \draw[dotted]  (-2,0.3)  --  (-0.3,2);

  \draw[dotted]  (2,1.5)  --  (1.5,2);
  \draw[dotted]  (2,0.9)  --  (0.9,2);
  \draw[dotted]  (2,0.3)  --  (0.3,2);

  \draw[dotted]  (-2,-1.5)  --  (-1.5,-2);
  \draw[dotted]  (-2,-0.9)  --  (-0.9,-2);
  \draw[dotted]  (-2,-0.3)  --  (-0.3,-2);

  \draw[dotted]  (2,-1.5)  --  (1.5,-2);
  \draw[dotted]  (2,-0.9)  --  (0.9,-2);
  \draw[dotted]  (2,-0.3)  --  (0.3,-2);

  \draw[dashed] (-4,0) -- (4,0);
  \draw[dashed] (-4,4) -- (4,4);
  \draw[dashed] (-4,-4) -- (-4,4);
  \draw[dashed] (4,-4) -- (4,4);
  \draw[dashed] (-4,-4) -- (4,-4);

  \draw[red] (-3.3,2.4) node{\scriptsize $\langle s_A|$};
  \draw[red] (-3.3,-2.4) node{\scriptsize $| s_A\rangle$};
  \draw[blue] (-2.2,3.5) node{\scriptsize $\langle s_B|$};
  \draw[blue] (-2.2,-3.5) node{\scriptsize $| s_B\rangle$};

  \draw[red] (3.3,-2.4) node{\scriptsize $\langle s_A|$};
  \draw[red] (3.3,2.4) node{\scriptsize $| s_A\rangle$};
  \draw[blue] (2.2,-3.5) node{\scriptsize $\langle s_B|$};
  \draw[blue] (2.2,3.5) node{\scriptsize $| s_B\rangle$};
\end{tikzpicture}\ \ \ \ \approx\ \ \ \ \begin{tikzpicture}[thick,scale = 0.45,baseline={([yshift=-4pt]current bounding box.center)}]
 \draw[red] (-2,-1.5) arc(0:-180:0.5 and 0.5);
 \draw[red] (-2,1.5) arc(0:180:0.5 and 0.5);
 \draw[red] (2,-1.5) arc(-180:0:0.5 and 0.5);
 \draw[red] (2,1.5) arc(180:0:0.5 and 0.5);
 \draw[blue] (-1.5,2) arc(-90:-270:0.5 and 0.5);
 \draw[blue] (-1.5,-2) arc(90:270:0.5 and 0.5);
 \draw[blue] (1.5,2) arc(-90:90:0.5 and 0.5);
 \draw[blue] (1.5,-2) arc(90:-90:0.5 and 0.5);

\filldraw  (-2.5,2) circle (1.5pt) node[left]{\scriptsize $ $};
\filldraw  (-2.5,-2) circle (1.5pt) node[left]{\scriptsize $ $};
\filldraw  (2.5,2) circle (1.5pt) node[left]{\scriptsize $ $};
\filldraw  (2.5,-2) circle (1.5pt) node[left]{\scriptsize $ $};

\filldraw  (-2,2.5) circle (1.5pt) node[left]{\scriptsize $ $};
\filldraw  (-2,-2.5) circle (1.5pt) node[left]{\scriptsize $ $};
\filldraw  (2,2.5) circle (1.5pt) node[left]{\scriptsize $ $};
\filldraw  (2,-2.5) circle (1.5pt) node[left]{\scriptsize $ $};

 \draw[red]  (-2,-1.5)  --  (-2,1.5);
 \draw[red]  (-3,-1.5)  --  (-3,1.5);
 \draw[red]  (2,-1.5)  --  (2,1.5);
 \draw[red]  (3,-1.5)  --  (3,1.5);
 
 \draw[blue]  (-1.5,-2)  --  (1.5,-2);
 \draw[blue]  (-1.5,-3)  --  (1.5,-3);
 \draw[blue]  (-1.5,2)  --  (1.5,2);
 \draw[blue]  (-1.5,3)  --  (1.5,3);

 \draw[red] (-2.5,0) node{$A$};
 \draw[blue] (0,2.5) node{$B$};
 \draw[red] (2.5,0) node{$A$};
 \draw[blue] (0,-2.5) node{$B$};

  \draw[dotted]  (-3,1.5)  --  (-2,1.5);
  \draw[dotted]  (-3,0.9)  --  (-2,0.9);
  \draw[dotted]  (-3,0.3)  --  (-2,0.3);
  \draw[dotted]  (-3,-1.5)  --  (-2,-1.5);
  \draw[dotted]  (-3,-0.9)  --  (-2,-0.9);
  \draw[dotted]  (-3,-0.3)  --  (-2,-0.3);

  \draw[dotted]  (3,1.5)  --  (2,1.5);
  \draw[dotted]  (3,0.9)  --  (2,0.9);
  \draw[dotted]  (3,0.3)  --  (2,0.3);
  \draw[dotted]  (3,-1.5)  --  (2,-1.5);
  \draw[dotted]  (3,-0.9)  --  (2,-0.9);
  \draw[dotted]  (3,-0.3)  --  (2,-0.3);

  \draw[dotted]  (1.5,3)  --  (1.5,2);
  \draw[dotted]  (0.9,3)  --  (0.9,2);
  \draw[dotted]  (0.3,3)  --  (0.3,2);
  \draw[dotted]  (-1.5,3)  --  (-1.5,2);
  \draw[dotted]  (-0.9,3)  --  (-0.9,2);
  \draw[dotted]  (-0.3,3)  --  (-0.3,2);

  \draw[dotted]  (1.5,-3)  --  (1.5,-2);
  \draw[dotted]  (0.9,-3)  --  (0.9,-2);
  \draw[dotted]  (0.3,-3)  --  (0.3,-2);
  \draw[dotted]  (-1.5,-3)  --  (-1.5,-2);
  \draw[dotted]  (-0.9,-3)  --  (-0.9,-2);
  \draw[dotted]  (-0.3,-3)  --  (-0.3,-2);

  \draw[dotted]  (-2,1.5)  --  (-1.5,2);
  \draw[dotted]  (-2,0.9)  --  (-0.9,2);
  \draw[dotted]  (-2,0.3)  --  (-0.3,2);

  \draw[dotted]  (2,1.5)  --  (1.5,2);
  \draw[dotted]  (2,0.9)  --  (0.9,2);
  \draw[dotted]  (2,0.3)  --  (0.3,2);

  \draw[dotted]  (-2,-1.5)  --  (-1.5,-2);
  \draw[dotted]  (-2,-0.9)  --  (-0.9,-2);
  \draw[dotted]  (-2,-0.3)  --  (-0.3,-2);

  \draw[dotted]  (2,-1.5)  --  (1.5,-2);
  \draw[dotted]  (2,-0.9)  --  (0.9,-2);
  \draw[dotted]  (2,-0.3)  --  (0.3,-2);

  \draw[dashed] (-4,0) -- (4,0);
  \draw[dashed] (-4,4) -- (4,4);
  \draw[dashed] (-4,-4) -- (-4,4);
  \draw[dashed] (4,-4) -- (4,4);
  \draw[dashed] (-4,-4) -- (4,-4);

  \draw[red] (-3.3,2.4) node{\scriptsize $\langle s_A|$};
  \draw[red] (-3.3,-2.4) node{\scriptsize $| s_A'\rangle$};
  \draw[blue] (-2.2,3.5) node{\scriptsize $\langle s_B|$};
  \draw[blue] (-2.2,-3.5) node{\scriptsize $| s_B'\rangle$};

  \draw[red] (3.3,-2.4) node{\scriptsize $\langle s_A'|$};
  \draw[red] (3.3,2.4) node{\scriptsize $| s_A\rangle$};
  \draw[blue] (2.2,-3.5) node{\scriptsize $\langle s_B'|$};
  \draw[blue] (2.2,3.5) node{\scriptsize $| s_B\rangle$};
\end{tikzpicture}
\end{equation}
 Here we drop the arrow for simplicity. Physically, this may be understood as follows: if we evolve for a long time, the reduced density matrix of subsystem $A$ always becomes thermal, and we can not distinguish different initial states $|\{s\}\rangle$ or $|\{s'\}\rangle$. If we sum over all possible boundary states, the contour becomes:
\begin{equation}
\sum_{s_A,s_A'}\sum_{s_B,s_B'}\begin{tikzpicture}[thick,scale = 0.45,baseline={([yshift=-0pt]current bounding box.center)}]
 \draw[red] (-2,-1.5) arc(0:-180:0.5 and 0.5);
 \draw[red] (-2,1.5) arc(0:180:0.5 and 0.5);
 \draw[red] (2,-1.5) arc(-180:0:0.5 and 0.5);
 \draw[red] (2,1.5) arc(180:0:0.5 and 0.5);
 \draw[blue] (-1.5,2) arc(-90:-270:0.5 and 0.5);
 \draw[blue] (-1.5,-2) arc(90:270:0.5 and 0.5);
 \draw[blue] (1.5,2) arc(-90:90:0.5 and 0.5);
 \draw[blue] (1.5,-2) arc(90:-90:0.5 and 0.5);

\filldraw  (-2.5,2) circle (1.5pt) node[left]{\scriptsize $ $};
\filldraw  (-2.5,-2) circle (1.5pt) node[left]{\scriptsize $ $};
\filldraw  (2.5,2) circle (1.5pt) node[left]{\scriptsize $ $};
\filldraw  (2.5,-2) circle (1.5pt) node[left]{\scriptsize $ $};

\filldraw  (-2,2.5) circle (1.5pt) node[left]{\scriptsize $ $};
\filldraw  (-2,-2.5) circle (1.5pt) node[left]{\scriptsize $ $};
\filldraw  (2,2.5) circle (1.5pt) node[left]{\scriptsize $ $};
\filldraw  (2,-2.5) circle (1.5pt) node[left]{\scriptsize $ $};

 \draw[red]  (-2,-1.5)  --  (-2,1.5);
 \draw[red]  (-3,-1.5)  --  (-3,1.5);
 \draw[red]  (2,-1.5)  --  (2,1.5);
 \draw[red]  (3,-1.5)  --  (3,1.5);
 
 \draw[blue]  (-1.5,-2)  --  (1.5,-2);
 \draw[blue]  (-1.5,-3)  --  (1.5,-3);
 \draw[blue]  (-1.5,2)  --  (1.5,2);
 \draw[blue]  (-1.5,3)  --  (1.5,3);

 \draw[red] (-2.5,0) node{$A$};
 \draw[blue] (0,2.5) node{$B$};
 \draw[red] (2.5,0) node{$A$};
 \draw[blue] (0,-2.5) node{$B$};

  \draw[dotted]  (-3,1.5)  --  (-2,1.5);
  \draw[dotted]  (-3,0.9)  --  (-2,0.9);
  \draw[dotted]  (-3,0.3)  --  (-2,0.3);
  \draw[dotted]  (-3,-1.5)  --  (-2,-1.5);
  \draw[dotted]  (-3,-0.9)  --  (-2,-0.9);
  \draw[dotted]  (-3,-0.3)  --  (-2,-0.3);

  \draw[dotted]  (3,1.5)  --  (2,1.5);
  \draw[dotted]  (3,0.9)  --  (2,0.9);
  \draw[dotted]  (3,0.3)  --  (2,0.3);
  \draw[dotted]  (3,-1.5)  --  (2,-1.5);
  \draw[dotted]  (3,-0.9)  --  (2,-0.9);
  \draw[dotted]  (3,-0.3)  --  (2,-0.3);

  \draw[dotted]  (1.5,3)  --  (1.5,2);
  \draw[dotted]  (0.9,3)  --  (0.9,2);
  \draw[dotted]  (0.3,3)  --  (0.3,2);
  \draw[dotted]  (-1.5,3)  --  (-1.5,2);
  \draw[dotted]  (-0.9,3)  --  (-0.9,2);
  \draw[dotted]  (-0.3,3)  --  (-0.3,2);

  \draw[dotted]  (1.5,-3)  --  (1.5,-2);
  \draw[dotted]  (0.9,-3)  --  (0.9,-2);
  \draw[dotted]  (0.3,-3)  --  (0.3,-2);
  \draw[dotted]  (-1.5,-3)  --  (-1.5,-2);
  \draw[dotted]  (-0.9,-3)  --  (-0.9,-2);
  \draw[dotted]  (-0.3,-3)  --  (-0.3,-2);

  \draw[dotted]  (-2,1.5)  --  (-1.5,2);
  \draw[dotted]  (-2,0.9)  --  (-0.9,2);
  \draw[dotted]  (-2,0.3)  --  (-0.3,2);

  \draw[dotted]  (2,1.5)  --  (1.5,2);
  \draw[dotted]  (2,0.9)  --  (0.9,2);
  \draw[dotted]  (2,0.3)  --  (0.3,2);

  \draw[dotted]  (-2,-1.5)  --  (-1.5,-2);
  \draw[dotted]  (-2,-0.9)  --  (-0.9,-2);
  \draw[dotted]  (-2,-0.3)  --  (-0.3,-2);

  \draw[dotted]  (2,-1.5)  --  (1.5,-2);
  \draw[dotted]  (2,-0.9)  --  (0.9,-2);
  \draw[dotted]  (2,-0.3)  --  (0.3,-2);

  \draw[dashed] (-4,0) -- (4,0);
  \draw[dashed] (-4,4) -- (4,4);
  \draw[dashed] (-4,-4) -- (-4,4);
  \draw[dashed] (4,-4) -- (4,4);
  \draw[dashed] (-4,-4) -- (4,-4);

  \draw[red] (-3.3,2.4) node{\scriptsize $\langle s_A|$};
  \draw[red] (-3.3,-2.4) node{\scriptsize $| s_A'\rangle$};
  \draw[blue] (-2.2,3.5) node{\scriptsize $\langle s_B|$};
  \draw[blue] (-2.2,-3.5) node{\scriptsize $| s_B'\rangle$};

  \draw[red] (3.3,-2.4) node{\scriptsize $\langle s_A'|$};
  \draw[red] (3.3,2.4) node{\scriptsize $| s_A\rangle$};
  \draw[blue] (2.2,-3.5) node{\scriptsize $\langle s_B'|$};
  \draw[blue] (2.2,3.5) node{\scriptsize $| s_B\rangle$};
\end{tikzpicture}\ \ \ \ =\ \ \ \ \begin{tikzpicture}[scale = 0.9,baseline={([yshift=0pt]current bounding box.center)}]
   \draw[blue,thick] (0.6,-0.85) arc(0:-360:0.6 and 0.6);
   \draw[blue,thick] (0.6,0.45) arc(0:-360:0.6 and 0.6);
   \draw[red,thick] (1.3,-0.2) arc(0:-360:1.3 and 1.3);

        \draw[dotted,thick] (0.4,-0.4) -- (1.126,-0.875);

        \draw[dotted,thick] (0.5,-1.191) -- (0.65,-1.35);

        \draw[dotted,thick] (0,-1.44) -- (0,-1.475);

        \draw[dotted,thick] (-0.5,-1.191) -- (-0.65,-1.35);

        \draw[dotted,thick] (-0.4,-0.4) -- (-1.126,-0.875);

        \draw[dotted,thick] (0.4,0) -- (1.126,0.475);

        \draw[dotted,thick] (0.5,0.791) -- (0.65,0.95);

        \draw[dotted,thick] (0,1.04) -- (0,1.075);

        \draw[dotted,thick] (-0.5,0.791) -- (-0.65,0.95);

        \draw[dotted,thick] (-0.4,0) -- (-1.126,0.475);

       \draw (-1.5,-0.2) node{\scriptsize$\beta$};
       \draw (1.5,0.2) node{\scriptsize $2\beta$};
       \draw (1.5,-0.6) node{\scriptsize $0$};
       \draw[thick,red] (1.25,-0.2) -- (1.35,-0.2);

      \draw[blue] (0.,-0.85) node{$B$};
      \draw[blue] (0.,0.45) node{$B$};
       \draw[red] (1.1,-1.4) node{$A$};
\end{tikzpicture}
\end{equation}
Here we have used the relation
\begin{equation}
\sum_{s}\ \begin{tikzpicture}[thick,scale = 0.45,baseline={([yshift=-2pt]current bounding box.center)}]
 \draw (-0.7,0.5) arc(90:-90:0.5 and 0.5);
 \draw (2.1,-0.5) arc(-90:-270:0.5 and 0.5);
 \draw (-0.7,-0.5) -- (-1.5,-0.5);
  \draw (-0.7,0.5) -- (-1.5,0.5);

 \draw (2.1,-0.5) -- (2.9,-0.5);
  \draw (2.1,0.5) -- (2.9,0.5);
  
  \draw (0.7,0) node{\scriptsize $|s \rangle\langle s|$};
 
 \filldraw  (-0.2,0) circle (1.5pt) node[left]{\scriptsize $ $};
 \filldraw  (1.6,0) circle (1.5pt) node[left]{\scriptsize $ $};
\end{tikzpicture}\ \ \ \ =\ \ \ \ \begin{tikzpicture}[thick,scale = 0.45,baseline={([yshift=-2pt]current bounding box.center)}]
 \draw (-1.5,-0.5) -- (1.5,-0.5);
  \draw (-1.5,0.5) -- (1.5,0.5);
\end{tikzpicture}\ ,
\end{equation}
and we have merged the contour for $\chi_1^S$ and $\chi_2^S$ for both subsystem $S=A/B$. This is exactly the contour for computing the subsystem R\'enyi entropy $\mathcal{S}_{A,th}^{(2)}$ of a thermal ensemble \cite{new}. Similarly, if we exchange the role of $A$ and $B$ subsystem, we get another saddle point corresponds to $\mathcal{S}_{B,th}^{(2)}$.

Note that this existence of the first-order transition can be viewed as an analogy of the information paradox in more complicated set-ups \cite{penington2019entanglement,almheiri2019entropy,almheiri2019page,almheiri2019islands,almheiri2019replica,penington2019replica,penington2019entanglement,Yiming}: if we consider increasing $\lambda$ and time $t$ from $\lambda=0$ and $t=0$, we could follow the saddle of $e^{-\mathcal{S}_{A,th}^{(2)}}$ without facing any singularity till $\lambda=1$, which leads to an information paradox. The solution to the information paradox is also similar: at some time $t$, a new saddle-point appears which preserves the unitary. For setups with clear bulk description, this new saddle corresponds to a solution with islands \cite{penington2019entanglement,almheiri2019entropy,almheiri2019page,almheiri2019islands,almheiri2019replica,penington2019replica,penington2019entanglement}.

For $q=2$, the system shows qualitatively different behaviors. Since the state $|\{1\}\rangle$ is annihilated by $c_j$, $|\psi(t)\rangle$ is then annihilated by $c_j(\beta/2+i t)$. For long time $t$, we expect this to be a random superposition $c_j(i\infty)=\sum_iO_{ji}\chi_i$ with some parameter $O_{ji}$. An remarkable observation is that this is just the ground state of SYK$_2$ model, and the entanglement entropy $\mathcal{S}_A^{(2)}$ is again given in \cite{new}. As shown in Figure \ref{fig3} (b), the arguments works even for $\beta=0$, which means the system would not thermalize to an infinite temperature ensemble $\rho=2^{-\frac{N}{2}} I$, as expected for non-interacting systems.

\begin{figure}[t]
  \center
  \includegraphics[width=1\columnwidth]{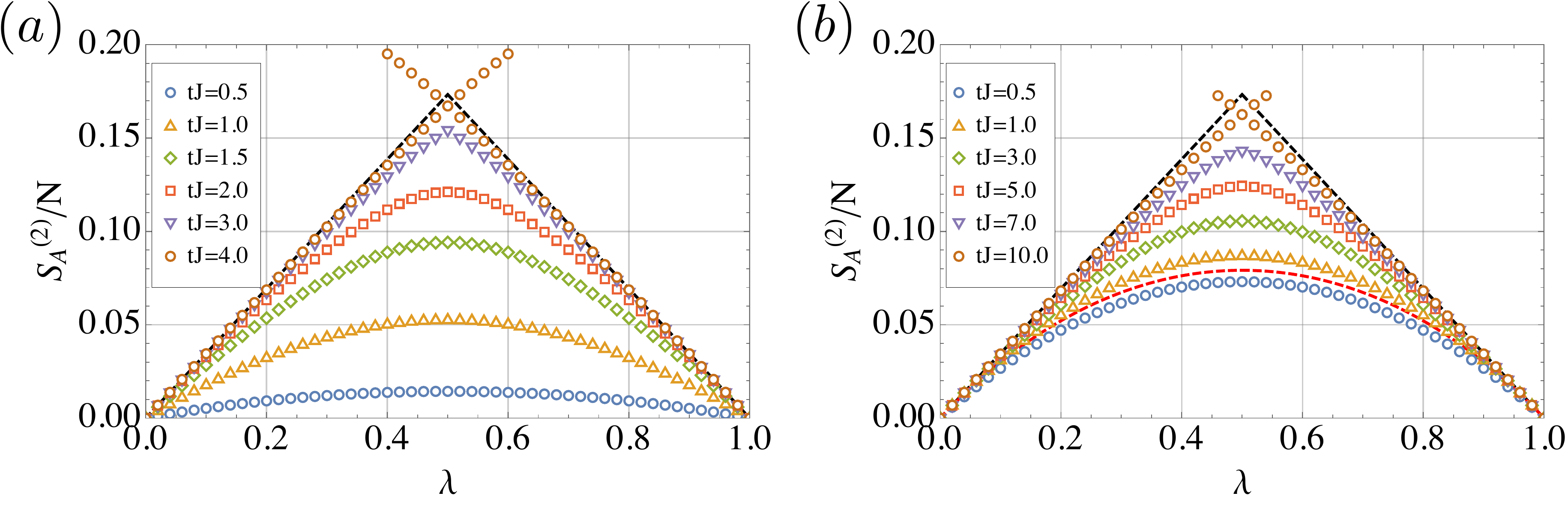}
  \caption{(a). The dynamics of the R\'enyi entanglement entropy $\mathcal{S}_A^{(2)}(t)$ as a function of subsystem size $\lambda$ with $V/J=1/2$ and $\beta=0$. The black dashed line is the maximal entropy: $\mathcal{S}_A^{(2)}(\lambda)/N=x \log(2)/2 $ with $x=\text{min}\{\lambda,1-\lambda\}$. (b). The dynamics of the R\'enyi entanglement entropy $\mathcal{S}_A^{(2)}(t)$ as a function of subsystem size $\lambda$ with $V/J=2$ and $\beta=0$. The black dashed line is the maximal entropy $\mathcal{S}_A^{(2)}(\lambda)/N=x \log(2)/2 $ with $x=\text{min}\{\lambda,1-\lambda\}$. The black dashed line is the analytical formula for the SYK$_2$ ground state \cite{new}. }\label{fig4}
 \end{figure}

We further consider adding SYK$_4$ random interaction term to the SYK$_2$ random hopping model \cite{banerjee2017solvable,chen2017competition,song2017strongly}. The Hamiltonian reads
\begin{equation}
H_V=\frac{1}{4!}\sum_{ijkl}J^{(4)}_{ijkl}\chi_{i}\chi_{j}\chi_{k}\chi_{l}+\frac{1}{2}\sum_{ij}iV^{(2)}_{ij}\chi_{i}\chi_{j}.
\end{equation}
Here to avoid possible confusion, we have changed the notion of the random hopping parameters to be $V^{(2)}_{ij}$ with variance $\overline{(V^{(2)}_{ij})^2}=V^2/N$. Near the SYK$_4$ fixed point, the random hopping term is relevant. As a result, the system is always a non-Fermi liquid for at low temperature $T\ll V^2/J$. At finite temperature, there is a crossover between the SYK$_2$ and the SYK$_4$ fixed points \cite{chen2017competition}.

We could similarly define the KM pure states for this model and study the entanglement dynamics with minor modification of \eqref{SDKMt}:
\begin{equation}
\begin{aligned}
\Sigma^A&=\Sigma^B=\left[J^2(\lambda G^A+(1-\lambda)G^B)^3+V^2(\lambda G^A+(1-\lambda)G^B)\right]\tilde{P}^{(2)}.
\end{aligned}
\end{equation}

The evolution of $\mathcal{S}_A^{(2)}$ for different $V/J$ is shown in Figure \ref{fig4}. We find that the entanglement show different behaviors for $V \lesssim J$ and $V \gtrsim J$. For $V \lesssim J$ the behavior is basically the same as the SYK$_4$ case in Figure \ref{fig3} (a), as shown in Figure \ref{fig4} (a). The reason is that the random hopping term plays an role only in the long time limit $t\sim J/V^2$, when the entanglement has already been built up. 

\begin{figure}[t]
  \center
  \includegraphics[width=1\columnwidth]{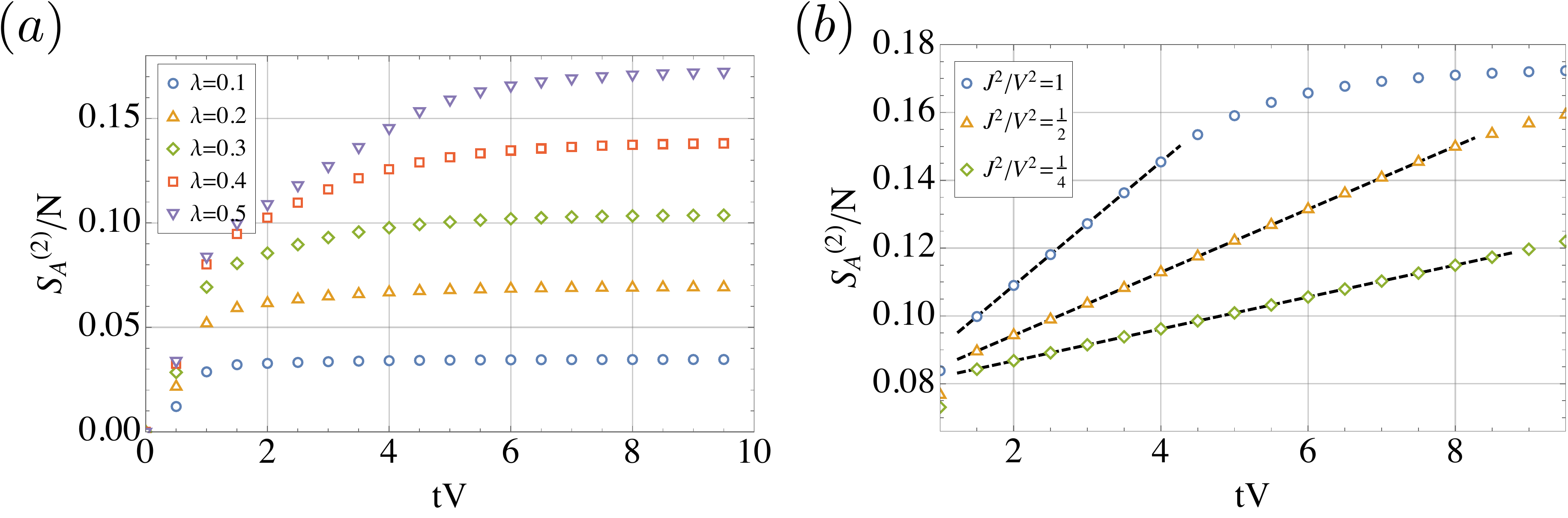}
  \caption{(a). The R\'enyi entanglement entropy $\mathcal{S}_A^{(2)}(t)$ as a function of time $tV$ for different subsystem size $\lambda$. Here we take $\beta=0$ and $V=J$. (a). The R\'enyi entanglement entropy $\mathcal{S}_A^{(2)}(t)$ as a function of time $tV$ for different subsystem size $V/J$. Here we fix $\beta=0$ and $\lambda=1/2$. The black dashed lines are linear fits for the linear growth region.}\label{fig5}
 \end{figure}

This is to be compared with the $V \gtrsim J$ case shown in Figure \ref{fig4} (b). In this case, the system builds up the entanglement in two steps. The short time behavior is governed by the random hopping term, leading to a $\mathcal{S}_A^{(2)}$ close to the SYK$_2$ ground state. Then the entanglement continues to increase linearly but with much slower speed until the system thermalizes. This can also been seen from a plot for $\mathcal{S}_A^{(2)}(t)$ with different time $t$. As shown in Figure \ref{fig5} (a), the entanglement entropy with $\lambda=1/2$ and $V=J$ increases rapidly for $tV<1$, while it increases linearly with a smaller slope for $tV>1$ until its saturation. The slope of the linear growth is proportional to $J^2/V$, which can be seen from Figure \ref{fig5} (b). This can be understood as a perturbative calculation near the $J=0$ solution, similar to the short-time behavior in \cite{Yiming,gu2017spread}.

\section{Conclusion}\label{sec6}

In this work, we study the entanglement R\'enyi entropy of Kourkoulou-Maldacena pure states of the SYK model, including its generalizations. We use the path-integral approach which gives the exact entanglement entropy in the large-$N$ limit. 

At low energy density, we find a first-order transition for the entanglement entropy for the SYK$_4$ model when tuning the subsystem size $\lambda$. This is different compared to the $q=2$ case where the entropy is a smooth function. Similar behaviors exist if we consider long real-time evolution. The first-order transition is from the existence of two different saddle points, which corresponds to the thermal R\'enyi entropy of reduced density matrices for different subsystems. We further consider adding small SYK$_4$ random interaction to the SYK$_2$ case, leading to the slow linear growth of entanglement entropy in the intermediate time regime.

{\bf \noindent Acknowledgement.} We thank Xiao Chen, Yingfei Gu, Chunxiao Liu for discussion. PZ acknowledges support from the Walter Burke Institute for Theoretical Physics at Caltech.

\bibliography{ref.bib}

\end{document}